\documentclass[english,pre,a4paper,aps,reprint]{revtex4-2}
\usepackage[T1]{fontenc}
\usepackage{amsmath}
\usepackage{amssymb}
\usepackage{graphicx}
\usepackage{babel}
\begin{document}
\title{Characterising quantum measurement through environmental stochastic
entropy production in a two spin 1/2 system}
\author{Sophia M. Walls, Adam Bloss and Ian J. Ford}
\address{Department of Physics and Astronomy, University College London, Gower
Street, London, WC1E 6BT, United Kingdom}
\email{Correspondence to: sophia.walls.17@ucl.ac.uk}

\selectlanguage{english}%
\begin{abstract}
Quantum state diffusion is a framework within which measurement may
be described as the continuous and gradual collapse of a quantum system
to an eigenstate as a result of interaction with its environment.
The irreversible nature of the quantum trajectories that arise may
be characterised by the environmental stochastic entropy production
associated with the measurement. We consider a system of two spin
1/2 particles undergoing either single particle measurements or measurements
of the total $z$-spin component $S_{z}$.  The mean asymptotic rates
of environmental stochastic entropy production associated with collapse
can depend on the eigenstate of $S_{z}$ selected, and on the initial
state of the system, offering an additional avenue for characterising
quantum measurement. 
\end{abstract}
\maketitle

\section{Introduction}

Entropy is a quantifier of the subjective uncertainty of the state
of a system and its rate of production is a measure of the irreversibility
of a process. It can arise when a system interacts with an underspecified,
complex environment according to dynamics characterised at a coarse-grained
level \citep{ford2013,lebowitz1993a}. The modern understanding of
entropy production was influenced particularly by the development
of fluctuation theorems \citep{evans1993,evans1995,evans2002,crooks1999,jarzynski1997,carberry2004,esposito2010}.
Systems driven by external noise can be characterised by a \emph{stochastic}
entropy production that can be divided into a system stochastic entropy
production and an environmental stochastic entropy production, each
with particular behaviour in a variety of circumstances \citep{gisin1993a,kurchan1998,lebowitz1999,seifert2005,harris2007,spinney2012a,spinney2012,ford2012,horowitz2013,leggio2013,ford2015a}.The
application of such ideas to quantum systems has contributed to the
field of quantum thermodynamics, from which technologies such as quantum
heat and measurement engines and refrigerators have emerged \citep{kosloff2014,solfanelli2023,jordan2020}.

Stochastic entropy production in the processing of quantum systems
has recently been explored, including studies of a two-level bosonic
system undergoing time dependent coupling to a harmonic environment,
an open three-level quantum system in a non-equilibrium steady state,
and entropy production associated with the quantum Zeno and quantum
anti-Zeno effects \citep{matos2022a,clarke2024,dexter2025,walls2024a,kinikar2023}.

The main focus of this paper is to consider the stochastic entropy
production when a system of two spin 1/2 particles is subjected to
measurement, extending work by Clarke and Ford\textit{ }on the stochastic
entropy production associated with measurements of a single spin 1/2
particle \citep{clarke2024}.  New features emerge, for example requiring
us to compute stochastic entropy production within a framework developed
to handle singular diffusion matrices \citep{dexter2025}.

Since average system evolution obscures the irreversibility of measurement,
we focus on single evolution pathways or quantum trajectories. A variety
of methods for generating quantum trajectories exist such as stochastic
Schr\"{o}dinger equations, hierarchical equations of motion, and quantum
jump trajectories comprised of piecewise deterministic dynamics interrupted
by stochastic discontinuities \citep{gambetta2003a,christie2022,carballeira2021,devega2017,horowitz2013,li2022b,gardiner1992,kloeden1992,breuer2004a,wiseman1993a}.
The Liouville-von Neumann equation offers an alternative approach
but does not produce physically meaningful trajectories since the
evolution does not preserve the unit trace \citep{stockburger2004,matos2022a}.
Many methods for generating quantum trajectories arise from unravellings
of the Lindblad master equation \citep{lindblad1976a}. A more recent
development has been jump-time unravellings, whereby quantum trajectories
are averaged at points where quantum jumps emerge \citep{gneiting2021}.

Projective measurements on quantum systems are discontinuous and instantaneous,
resulting in difficulties when attempting to find an associated stochastic
entropy production. In contrast, quantum state diffusion (QSD) produces
continuous, stochastic quantum trajectories without jumps and it is
this framework that we use to generate trajectories \citep{wiseman1993a,wiseman1996a,diosi1998a,gisin1993a,gisin1996}.
The quantum system is considered to interact with an environment which
causes it to exhibit a diffusive evolution. A large environment requires
knowledge of many degrees of freedom, so its state can not be specified
and instead its influence on the system is taken to be stochastic.
The system's behaviour is therefore akin to that of a classical Brownian
particle. For the measurement process, we consider the environment
to act as a measurement apparatus whose interaction with the system
causes it to diffuse towards an eigenstate of the system observable
being measured.

The coupling between the system and the environment characterises
the strength of measurement. In the limit of infinitely strong coupling,
the system would seem to exhibit the conventional, discontinuous quantum
jumps, whilst at low coupling it undergoes weak measurement, resulting
in ongoing extraction of partial information about the system. The
evolution of the quantum system may in some circumstances then be
described by a stochastic differential equation (SDE), or an It\^{o}
process, featuring a noise term that represents the influence of the
environment on the system \citep{gisin1992b,gisin1997,wiseman1993a,wiseman1996a,percival1998a}.

Utilising QSD, we generate quantum trajectories for a system of two
spin 1/2 particles undergoing either measurements of single particle
spin components, or measurements of a component of the total spin
of the two particles. We calculate the rate of environmental stochastic
entropy production associated with each measurement process and compare
the mean asymptotic rates of production for approach towards different
eigenstates and for different initial conditions. 

The plan for the paper is as follows: Section \ref{sec:Quantum-state-diffusion}
introduces the framework of QSD with Section \ref{sec:Environmental-entropy-production}
describing the key concepts relating to stochastic entropy production
in a situation characterised by singular diffusion matrices. Section
\ref{sec:local_mes} considers single particle spin measurements of
the particles, whilst Section \ref{sec:global_mes} extends this to
measurements of total spin. Conclusions are given in Section \ref{sec:Conclusions}.

\section{Quantum state diffusion \label{sec:Quantum-state-diffusion}}

In QSD, measurement is modelled as an interaction between a system
and its environment that causes the system to evolve continuously
and stochastically towards an eigenstate of the measured operator
\citep{percival1998a,gisin1992b,gisin1997}. The evolution of $\rho$,
the reduced density matrix of the system, in a time-step $dt$, is
described in terms of Kraus operators $M_{j}$ \citep{tong2006a,nielsen2002a}:

\begin{equation}
\rho(t+dt)=\rho(t)+d\rho=\frac{M_{j}\rho(t)M_{j}^{\dagger}}{{\rm Tr}(M_{j}\rho(t)M_{j}^{\dagger})},\label{eq:single_kraus-1}
\end{equation}
and such a transition is considered to occur with probability $p_{j}={\rm Tr}(M_{j}\rho(t)M_{j}^{\dagger})$
\citep{walls2024a}.

In order to generate continuous trajectories, Kraus operators that
differ incrementally from a multiple of the identity are used:

\begin{equation}
M_{j}\equiv M_{k\pm}=\frac{1}{\sqrt{2}}(\mathbb{I}+A_{k\pm}),\label{eq:1+A}
\end{equation}
with
\begin{equation}
A_{k\pm}=-iH_{s}dt-\frac{1}{2}L_{k}^{\dagger}L_{k}dt\pm L_{k}\sqrt{dt},\label{eq:Ak}
\end{equation}
where the system Hamiltonian is denoted by $H_{s}$, Lindblad operators
by $L_{k}$ and each Lindblad channel is associated with two Kraus
operators labelled by $\pm$ \citep{jacobs2014,gross2018a,matos2022a,walls2024a,walls2024,clarke2024}.
These Kraus operators satisfy the completeness condition $\sum_{j}M_{j}^{\dagger}M_{j}=\mathbb{I}$.
Since the future state of the system depends only on its current state
$\rho(t)$, the evolution is Markovian.

A map with a Kraus operator of the form given in Eqs. (\ref{eq:1+A})
and (\ref{eq:Ak}) has been demonstrated to preserve positivity \citep{walls2024a}.
A sequence of such transitions then yields a stochastic trajectory
which corresponds to the solution to the following SDE:

\begin{align}
d\rho & =-i[H_{s},\rho]dt+\sum_{k}\Big((L_{k}\rho L_{k}^{\dagger}-\frac{1}{2}\{L_{k}^{\dagger}L_{k},\rho\})dt\nonumber \\
 & \qquad+\left(\rho L_{k}^{\dagger}+L_{k}\rho-{\rm Tr}[\rho(L_{k}+L_{k}^{\dagger})]\rho\right)dW_{k}\Big),\label{eq:SME-1}
\end{align}
where $\hbar$ has been set to unity. The evolution of $\rho$ is
thus an It\^o process with Wiener increments $dW_{k}$, the index
$k$ denoting the Lindblad operator \citep{gisin1997}. The stochasticity
arises from system-environment interactions: in the absence of such
interactions with the environment the evolution of $\rho$ reduces
to the von Neumann equation.

It has been demonstrated that the average evolution over all possible
system transitions and over all possible system states satisfies a
Lindblad master equation \citep{matos2022a,walls2024,walls2024a}
\begin{equation}
d\bar{\rho}=-i[H_{s},\bar{\rho}]dt+\sum_{k}\left(L_{k}\bar{\rho}L_{k}^{\dagger}-\frac{1}{2}\{L_{k}^{\dagger}L_{k},\bar{\rho}\}\right)dt,\label{eq:lindblad eq}
\end{equation}
for an ensemble averaged reduced density matrix $\bar{\rho}(t)$.
We employ an overbar as a reminder of the averaging, though this is
not the usual notation in the literature. In spite of the invariance
of the right hand side of Eq. (\ref{eq:lindblad eq}) under unitary
transformations amongst the Kraus operators, the requirement that
the underlying stochastic trajectories are continuous means that the
Kraus operators are uniquely given by Eqs. (\ref{eq:1+A}) and (\ref{eq:Ak})
in this case.

\section{Environmental stochastic entropy production \label{sec:Environmental-entropy-production}}

\subsection{Dynamics of entropy production}

The von Neumann entropy is an extension of the Gibbs entropy to quantum
systems and provides a measure of the uncertainty over the possible
outcomes associated with a projective measurement of a system starting
in a given state. It is defined as $-\sum_{i}P_{i}\text{ln}P_{i}$
where $P_{i}$ is the probability of projection onto an eigenstate
$i$ of the measured observable \citep{nielsen2002a}. It can also
be written as $-{\rm Tr}\rho\ln\rho$ in terms of the density matrix.

However, we wish to characterise the irreversibility of continuous
single realisations of system behaviour in its Hilbert space, not
projective measurement, so the von Neumann entropy is not suitable.
It is stochastic entropy production that provides this: a change in
subjective uncertainty of the quantum state of the system and its
environment arising from our inability to make precise predictions
when the state of the environment and its influence on the system
is underspecified \citep{seifert2005,spinney2012}. Whilst the von
Neumann entropy has an upper limit of $\text{ln}2$ for a spin $1/2$
system, the stochastic entropy production is unbounded since it concerns
the uncertainty of adoption over a continuum of possible states.

To be specific, the (total) stochastic entropy production $\Delta s_{{\rm tot}}$
associated with a trajectory is defined by the ratio of probabilities
of the (forward) trajectory and the reverse sequence of events driven
by a reverse protocol of external forces:

\begin{equation}
\Delta s_{{\rm tot}}=\text{ln}\left(\frac{\text{prob}(\text{forward trajectory)}}{\text{prob(\text{reverse trajectory)}}}\right).
\end{equation}
This can be separated into two contributions, system $\Delta s_{{\rm sys}}$
and environmental $\Delta s_{{\rm env}}$. The former depends on the
probability density function (pdf) over the space of system microstates
while the latter may be derived from the stochastic dynamics that
govern the system evolution.

We employ the following framework. A set of $N$ coordinates $\boldsymbol{x}\equiv(x_{1},x_{2},\cdots,x_{N})$
specifies the configuration of a system, and their evolution is modelled
using Markovian stochastic differential equations, or It\^{o} processes,
of the form

\begin{equation}
dx_{i}=A_{i}(\boldsymbol{x})dt+\sum_{j=1}^{M}B_{ij}(\boldsymbol{x})dW_{j},\label{eq:a100-1}
\end{equation}
where the $dW_{j}$ are $M$ independent Wiener increments. Defining
an $N\times N$ diffusion matrix $\boldsymbol{D}(\boldsymbol{x})=\frac{1}{2}\boldsymbol{B}(\boldsymbol{x})\boldsymbol{B}(\boldsymbol{x})^{\mathsf{T}}$,
the Fokker-Planck equation for the pdf $p(\boldsymbol{x},t)$ is given
by
\begin{align}
\frac{\partial p}{\partial t} & =-\sum_{i}\frac{\partial}{\partial x_{i}}\left(A_{i}p\right)+\sum_{ij}\frac{\partial}{\partial x_{i}\partial x_{j}}\left(D_{ij}p\right),\label{eq:fpe}\\
 & =-\sum_{i}\frac{\partial}{\partial x_{i}}\left(C_{i}p-D_{ij}\frac{\partial p}{\partial x_{j}}\right),\nonumber 
\end{align}
where 
\begin{equation}
C_{i}=A_{i}-\sum_{j}\frac{\partial D_{ij}}{\partial x_{j}}.\label{eq:3}
\end{equation}

The system stochastic entropy production is given by $d\Delta s_{{\rm sys}}=-d\ln p(\boldsymbol{x},t)$
but for simplicity we focus our attention only on the environmental
stochastic entropy production. We consider coordinates with \emph{even
parity} under time reversal. The evolution of $\Delta s_{{\rm env}}$
is then governed by the SDE \citep{lebowitz1999,spinney2012a,spinney2012}
\begin{align}
d\Delta s_{\text{env}} & =\sum_{ij}D_{ij}^{-1}C_{j}dx_{i}+\sum_{ijk}D_{ij}\frac{\partial\left(D_{ik}^{-1}C_{k}\right)}{\partial x_{j}}dt,\nonumber \\
 & =\sum_{ij}D_{ij}^{-1}C_{j}\circ dx_{i}.\label{eq: senvbig}
\end{align}
A general expression suitable for systems with odd as well as even
parity coordinates is given in Appendix \ref{sec:Appendix-A}.

\subsection{Procedure for singular diffusion matrices \label{subsec:Singular-Diffusion-Matrices}}

Singular diffusion matrices arise for systems where there are more
coupled It\^{o} processes than independent Wiener noise increments.
At points in the coordinate space of the system there are directions
for which the rate of diffusion is zero. These directions are parallel
to eigenvectors of the diffusion matrix with zero eigenvalues, denoted
null eigenvectors. The singularity in the diffusion matrix can be
attributed to the existence of a time independent or a deterministically
evolving function of the variables $x_{i}$ \citep{dexter2025}. The
procedure for dealing with singular diffusion matrices involves focussing
on a sub-set of the total phase space spanned by $M$ coordinates,
which we denote \emph{dynamical} variables. The remaining $L=N-M$
coordinates will be regarded as \emph{spectator} variables.

Such a distinction between dynamical and spectator variables is arbitrary
since the total entropy production does not depend on which variable
is chosen to be dynamical, but a careful choice can greatly simplify
the computation. The reduced diffusion matrix defined in the co-ordinate
sub-space with dimensions $M\times M$ is non-singular so analysis
using Eq. (\ref{eq: senvbig}) can proceed. The spectator variables
nevertheless still play a role in computing the entropy production.
The main implication is that the derivatives $dD_{ij}/dx_{m}$ are
replaced by a new expression involving derivatives with respect to
dynamical variables $x_{m}$ and the spectator variables $x_{l}$:

\begin{equation}
\frac{dD_{ij}}{dx_{m}}=\frac{\partial D_{ij}}{\partial x_{m}}+\sum_{l}\frac{\partial D_{ij}}{\partial x_{l}}R_{lm},\label{eq:corr_terms}
\end{equation}
where $R_{lm}$ is a matrix formed of components of the null eigenvectors
of the full $N\times N$ diffusion matrix. See Appendix \ref{sec:Appendix-B}
and \citep{dexter2025} for further details.

\section{Two Spin 1/2 particles undergoing separate measurements \label{sec:local_mes}}

\subsection{System specification and dynamics}

The density matrix describing a system of two spin 1/2 particles may
be constructed as follows:

\begin{equation}
\rho=\frac{1}{4}(\mathbb{I}+\boldsymbol{s}\cdot\boldsymbol{\Sigma}),\label{eq:rho_ent_spins}
\end{equation}
using the 15 component coherence vector $\boldsymbol{s}=(s_{1},s_{2},\cdots,s_{15})$
and the vector $\boldsymbol{\Sigma}$ of generators $\Sigma_{m}$,
with $m=1,15$, built using Pauli matrices: details are given in Appendix
\ref{sec:Appendix-C}. The density matrix may then be written as:\begin{widetext}
\begin{equation}
\rho=\frac{1}{4}\begin{pmatrix}1+s_{3}+s_{12}+s_{15} & -is_{2}+s_{13}-is_{14}+s_{1} & s_{4}+s_{7}-is_{8}-io & s_{5}-is_{6}-is_{9}-s_{10}\\
is_{2}+s_{13}+is_{14}+s_{1} & 1-s_{3}+s_{12}-s_{15} & s_{5}+is_{6}-is_{9}+s_{10} & s_{4}-s_{7}-is_{8}+is_{11}\\
s_{4}+s_{y}+is_{8}+is_{11} & s_{5}-is_{6}+is_{9}+s_{10} & 1+s_{3}-s_{12}-s_{15} & -is_{2}-s_{13}+is_{14}+s_{1}\\
s_{5}+is_{6}+is_{9}-s_{10} & s_{4}-s_{7}+is_{8}-is_{11} & is_{2}-s_{13}-is_{14}+s_{1} & 1-s_{3}-s_{12}+s_{15}
\end{pmatrix}.\label{eq:ent_spin_density_matrix}
\end{equation}
\end{widetext}Stochastic trajectories of $\rho$ are generated from
SDEs for the components of the coherence vector, $s_{m}$, produced
from

\begin{equation}
ds_{m}={\rm Tr}(d\rho\Sigma_{m}),\label{eq:ent_spin_eqn_motion}
\end{equation}
where $d\rho$ is defined in Eq. (\ref{eq:SME-1}). We now consider
different scenarios of single particle measurements performed on the
two spin 1/2 system, corresponding to particular choices of Lindblad
operators.

\subsection{Case 1: $z$-spins of each particle}

We first consider simultaneous measurements of the $z$-component
of each of the two spins in the system. The system starts in the singlet
state $|\Psi^{-}\rangle=\frac{1}{\sqrt{2}}(|1\rangle_{z}|-1\rangle_{z}-|-1\rangle_{z}|1\rangle_{z})$,
where $\vert\pm1\rangle_{z}$ are the eigenstates of $z$-spin for
each particle. The Lindblad operators are expressed as $L_{1}=a_{1}\frac{1}{2}\sigma_{z,1}\otimes\mathbb{I}$
and $L_{2}=\mathbb{I}\otimes a_{2}\frac{1}{2}\sigma_{z,2}$, where
the scalar coefficients $a_{1}$ and $a_{2}$ represent couplings
between the system and the environment. Insertion into Eq. (\ref{eq:SME-1})
yields fifteen SDEs for the coherence vector components, given in
Appendix \ref{sec:Appendix-D}. The SDEs feature two noises, one for
each measurement in the system. The `expectation values' of the individual
$z$-spin components may be considered to be physical attributes of
the quantum state \citep{walls2024a,walls2024} and can be expressed
as:

\begin{align}
\langle S_{z,1}\rangle & =\text{Tr}\left(\left(\frac{1}{2}\sigma_{z,1}\otimes\mathbb{I}\right)\rho\right)=\frac{1}{2}s_{3}\nonumber \\
\langle S_{z,2}\rangle & =\text{Tr}\left(\left(\mathbb{I}\otimes\frac{1}{2}\sigma_{z,2}\right)\rho\right)=\frac{1}{2}s_{12}.
\end{align}
The SDEs take the form: $dx_{i}=A_{i}dt+B_{i1}dW_{1}+B_{i2}dW_{2}$
and the diffusion matrix elements are $D_{ij}=\frac{1}{2}(B{}_{i1}B_{j1}+B_{i2}B_{j2})$.
There is only one non-zero eigenvalue, so we use the method outlined
in section \ref{subsec:Singular-Diffusion-Matrices}, choosing $s_{12}$
as the dynamical variable since its associated SDE has $A_{12}=0$,
which simplifies the calculation. The reduced (scalar) diffusion coefficient,
for $a_{1}=a_{2}=1$, is written

\begin{equation}
D=\frac{1}{2}(-1+s_{12})^{2}+\frac{1}{2}(s_{3}s_{12}-s_{15})^{2},\label{eq:15a}
\end{equation}
and Eq. (\ref{eq: senvbig}) reduces to

\begin{equation}
d\Delta s_{{\rm env}}=-\frac{1}{D}\frac{dD}{ds_{12}}ds_{12}-\frac{d^{2}D}{ds_{12}^{2}}dt+\frac{1}{D}\left(\frac{dD}{ds_{12}}\right)^{2}dt,\label{eq:1d_stoch_ent-1}
\end{equation}
with the derivative $dD/ds_{12}$ given by 
\begin{equation}
\frac{dD}{ds_{12}}=\frac{\partial D}{\partial s_{12}}+\sum_{l\ne12}\frac{\partial D}{\partial s_{l}}R_{l,12}.\label{eq:18}
\end{equation}
The diffusion coefficient only depends on the variables $s_{12},s_{3}$
and $s_{15}$, so the relevant spectator variables are $s_{3}$ and
$s_{15}$. Starting in the singlet state $|\Psi^{-}\rangle=\frac{1}{\sqrt{2}}(|1\rangle_{z}|-1\rangle_{z}-|-1\rangle_{z}|1\rangle_{z})$,
we find that $R_{3,12}=-1$ and $R_{15,12}=0$ throughout. Hence
the evolution of the environmental stochastic entropy production may
be written as: \begin{widetext}

\begin{align}
d\Delta s_{{\rm env}} & =\left(-(s_{3}^{2}-4s_{3}s_{12}+7s_{12}^{2}+2s_{15}-2)+\frac{2\left(2s_{3}(s_{3}s_{12}-s_{15})+4s_{12}(s_{12}^{2}-1)-2s_{12}(s_{3}s_{12}-s_{15})\right)^{2}}{(s_{12}^{2}-1)^{2}+(s_{3}s_{12}-s_{15})^{2}}\right)dt\nonumber \\
 & +\frac{2\left(s_{3}(s_{3}s_{12}-s_{15})+2s_{12}(s_{12}^{2}-1)-s_{12}(s_{3}s_{12}-s_{15})\right)\left(\left(s_{12}^{2}-1\right)dW_{1}+\left(s_{3}s_{12}-s_{15}\right)dW_{2}\right)}{(s_{12}^{2}-1)^{2}+(s_{3}s_{12}-s_{15})^{2}}.\label{eq:z_z_ent_prod}
\end{align}

\end{widetext}

\subsection{Case 2: $z$-spin of particle 1 and $x$-spin of particle 2}

Next we consider the case where particle 1 undergoes a $z$-spin measurement
and particle 2 an $x$-spin measurement. Using Lindblad operators
$L_{1}=a_{1}\frac{1}{2}\sigma_{z,1}\otimes\mathbb{I}$ and $L_{2}=\mathbb{I}\otimes a_{2}\frac{1}{2}\sigma_{x,2}$
leads again to fifteen SDEs, shown in Appendix \ref{sec:Appendix-E}.
The SDEs contain two noises $dW_{1}$ and $dW_{2}$ relating to the
measurements on the first and second particle. The `expectation values'
of the $z$-spin of the first particle and the $x$-spin of the second
particle are:

\begin{align}
\langle S_{z,1}\rangle= & \text{Tr}(\frac{1}{2}\sigma_{z,1}\rho)=\frac{1}{2}s_{12}\nonumber \\
\langle S_{x,2}\rangle= & \text{Tr}(\frac{1}{2}\sigma_{x,2}\rho)=\frac{1}{2}s_{1}.
\end{align}
The diffusion matrix this time contains two non-zero eigenvalues,
implying two dynamical variables that we choose to be $s_{1}$ and
$s_{12}$. The reduced diffusion matrix in the basis of $s_{1}$ and
$s_{12}$ variables is \begin{widetext}

\begin{equation}
\boldsymbol{D}=\frac{1}{2}\begin{pmatrix}(-1+s_{1}^{2})^{2}+(s_{1}s_{12}-s_{13})^{2}, & (-2+s_{1}^{2}+s_{12}^{2})(s_{1}s_{12}-s_{13})\\
(-2+s_{1}^{2}+s_{12}^{2})(s_{1}s_{12}-s_{13}), & (-1+s_{12}^{2})^{2}+(s_{1}s_{12}-s_{13})^{2}
\end{pmatrix}.
\end{equation}
\end{widetext} Since $s_{1},s_{12}$ and $s_{13}$ form a closed
group of equations, $s_{13}$ is the only spectator variable we need
to consider. We can express the derivative of the diffusion matrix
inverse for use in Eq. (\ref{eq: senvbig}) using the following:

\begin{equation}
\begin{aligned}\frac{\partial(D^{-1}D)}{\partial x_{k}}=\frac{\partial D^{-1}}{\partial x_{k}}D+D^{-1}\frac{\partial D}{\partial x_{k}}=\frac{\partial\mathbb{I}}{\partial x_{k}}=0\end{aligned}
,
\end{equation}
and hence
\begin{equation}
\frac{\partial D^{-1}}{\partial x_{k}}=-D^{-1}\frac{\partial D}{\partial x_{k}}D^{-1}.\label{eq:39a}
\end{equation}
We do not present a full analytic expression for the environmental
stochastic entropy production due to its complexity.

\subsection{Environmental stochastic entropy production for Cases 1 and 2}

\begin{figure}
\begin{centering}
\includegraphics[width=1\columnwidth]{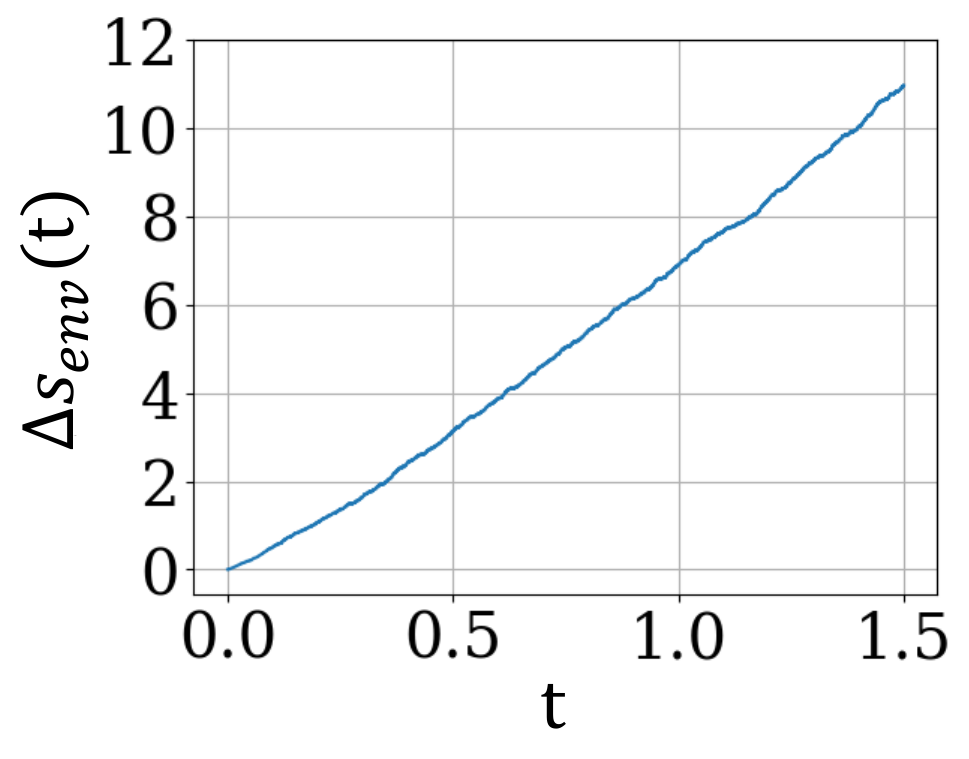}
\par\end{centering}
\caption{The mean environmental stochastic entropy production for Case 1:
two spin 1/2 particles each undergoing $z$-spin measurements starting
at $t=0$. Initial state: $|\Psi^{-}\rangle=\frac{1}{\sqrt{2}}(|1\rangle_{z}|-1\rangle_{z}-|-1\rangle_{z}|1\rangle_{z})$,
measurement strengths $a_{1}=a_{2}=1$, and an average over 326 trajectories
using time-step 0.0001.  \label{fig:z_z_mes_ent} }
\end{figure}

The mean environmental stochastic entropy production associated with
Case 1 is illustrated in Figure \ref{fig:z_z_mes_ent}. The mean rate
associated with measurements of the $z$-spin of each particle is
approximately double that arising from measurement of the $z$-spin
of a single particle \citep{clarke2024}. Since the particles start
in the singlet state, there are only two possible measurement outcomes:
$|1\rangle_{z}|-1\rangle_{z}$ and $|-1\rangle_{z}|1\rangle_{z}$,
such that $s_{3}=-s_{12}$ and asymptotically $s_{3}\to\pm1$. Furthermore,
it may be shown that $s_{15}$ remains equal to $-1$, and hence Eq.
(\ref{eq:z_z_ent_prod}) can be simplified to $d\Delta s_{{\rm env}}=4(s_{12}^{2}+1)dt+4s_{12}(dW_{1}-dW_{2})$.
This suggests an asymptotic mean rate of environmental stochastic
entropy production of $d\langle\Delta s_{{\rm env}}\rangle/dt\to8$,
which is reflected in the slope in Figure \ref{fig:z_z_mes_ent}.
Note that the mean rate of entropy production is considered to be
asymptotic when the entropy production exhibits a linear increase
with time and the system is close to an eigenstate.

\begin{figure}
\begin{centering}
\includegraphics[width=1\columnwidth]{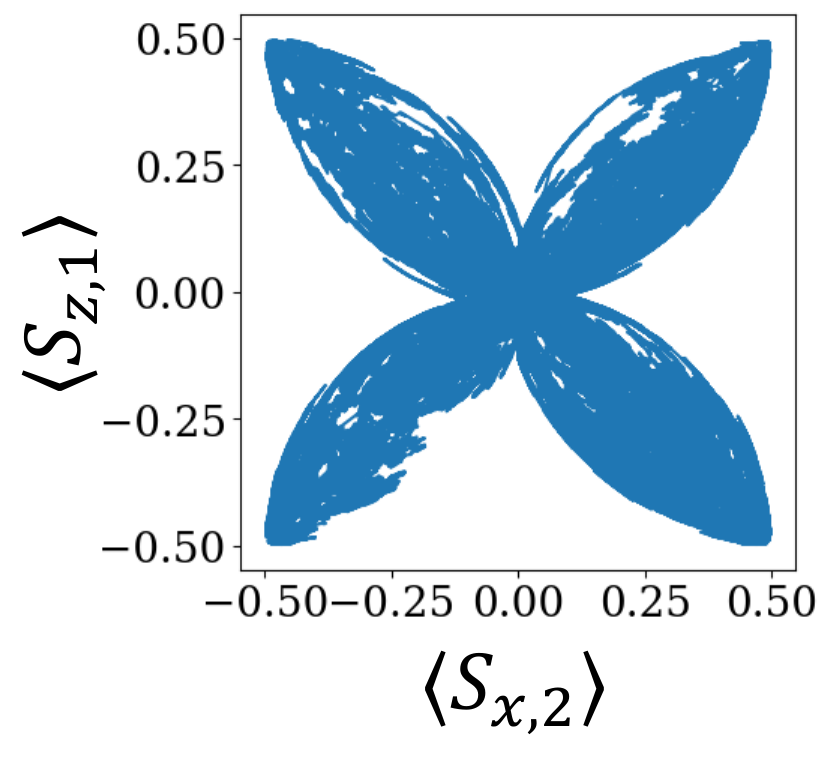}
\par\end{centering}
\caption{An illustration of 500 example stochastic trajectories in the space
of $\langle S_{x,2}\rangle$ and $\langle S_{z,1}\rangle$ for Case
2: two spin 1/2 particles where a $z$-spin measurement is performed
on particle 1 and an $x$-spin measurement is performed on particle
2. Trajectories begin at the state $|\Psi^{-}\rangle=\frac{1}{\sqrt{2}}(|1\rangle_{z}|-1\rangle_{z}-|-1\rangle_{z}|1\rangle_{z})$,
corresponding to $\langle S_{x,2}\rangle=\langle S_{z,1}\rangle=0$,
and proceed towards the tip of one of the `petals'. Measurement strengths
$a_{1}=a_{2}=1$ and time-step 0.0001. \label{fig:z_x_example_traj}}
\end{figure}

Figure \ref{fig:z_x_example_traj} illustrates a set of example trajectories
for Case 2: where a $z$-spin measurement is performed on particle
1 and an $x$-spin measurement is performed on particle 2. The initial
state is: $|\Psi^{-}\rangle=\frac{1}{\sqrt{2}}(|1\rangle_{z}|-1\rangle_{z}-|-1\rangle_{z}|1\rangle_{z})$,
which lies at $\langle S_{x,2}\rangle=\langle S_{z,1}\rangle=0$.
Particle 1 has two possible measurement outcomes: $|1\rangle_{z}$
and $|-1\rangle_{z}$ and similarly particle 2 has possible measurement
outcomes: $|1\rangle_{x}$ and $|-1\rangle_{x}$. Hence there are
four possible joint outcomes: $|1\rangle_{z}|1\rangle_{x},|1\rangle_{z}|-1\rangle_{x},|-1\rangle_{z}|1\rangle_{x}$
and $|-1\rangle_{z}|-1\rangle_{x}$ which lie at the tips of the petal
shape in Figure \ref{fig:z_x_example_traj}.

\begin{figure}
\begin{centering}
\includegraphics[width=1\columnwidth]{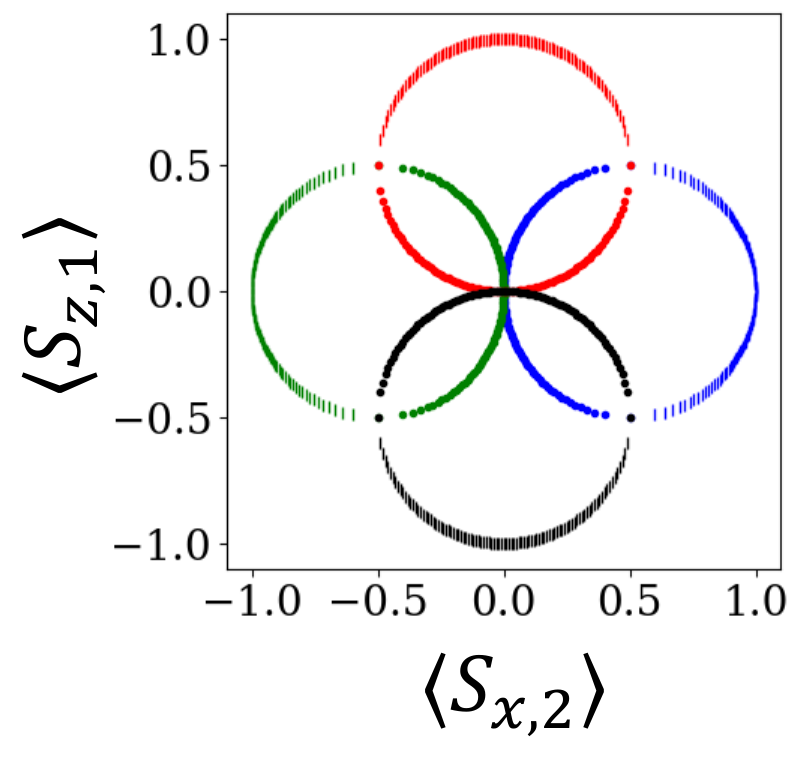}
\par\end{centering}
\caption{Projections onto the $\langle S_{x,2}\rangle$,$\langle S_{z,1}\rangle$
space of superpositions (with real amplitudes) of two of the four
measurement outcomes for Case 2. Blue: $|1\rangle_{z}|1\rangle_{x}\text{ and }|-1\rangle_{z}|1\rangle_{x}$;
green: $|-1\rangle_{z}|-1\rangle_{x}\text{ and }|1\rangle_{z}|-1\rangle_{x}$;
red: $|1\rangle_{z}|1\rangle_{x}\text{ and }|1\rangle_{z}|-1\rangle_{x}$;
black: $|-1\rangle_{z}|-1\rangle_{x}\text{ and }|-1\rangle_{z}|1\rangle_{x}$.
Points represented by dots signify states that are dynamically accessible
to the system under the specific measurement scheme. Points illustrated
by short vertical lines are not accessible but illustrate the broader
pattern. The dynamics constrain the stochastic trajectories to lie
in the overlap regions between the circles, forming the petal shapes
in Figure \ref{fig:z_x_example_traj}. \label{fig:petal_superpositions}}
\end{figure}

The pattern arises because the stochastic trajectories are constrained
to lie between loci defined by superpositions (with real amplitudes)
of pairs of measurement outcomes $|\pm1\rangle_{z}|\pm1\rangle_{x}$,
depicted in Figure \ref{fig:petal_superpositions}. The blue circles
illustrate a superposition of the $|1\rangle_{z}|1\rangle_{x}\text{ and }|-1\rangle_{z}|1\rangle_{x}$
measurement outcomes; green $|-1\rangle_{z}|-1\rangle_{x}\text{ and }|1\rangle_{z}|-1\rangle_{x}$;
red $|1\rangle_{z}|1\rangle_{x}\text{ and }|1\rangle_{z}|-1\rangle_{x}$;
and black $|-1\rangle_{z}|-1\rangle_{x}\text{ and }|-1\rangle_{z}|1\rangle_{x}$.
Points on the circles represented by short vertical lines in Figure
\ref{fig:petal_superpositions} cannot be reached by the dynamics
of the system because they concern values of $\langle S_{x,2}\rangle$
and $\langle S_{z,1}\rangle$ that are greater than $\pm\frac{1}{2}$.
In contrast, the points represented by dots are accessible.

\begin{figure}
\begin{centering}
\includegraphics[width=1\columnwidth]{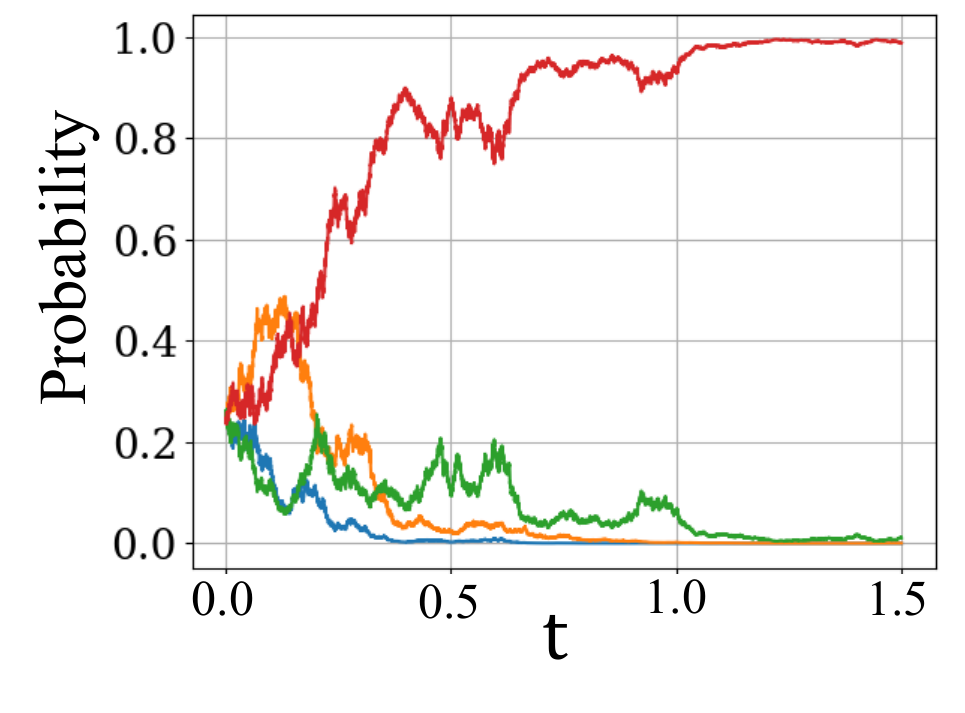}
\par\end{centering}
\caption{Example evolution of the probability amplitudes of the possible
measurement outcomes in Case 2: the two spin 1/2 system undergoing
a $z$-spin measurement on particle 1 and an $x$-spin measurement
on particle 2. Red: the probability amplitude of the $|1\rangle_{z}|1\rangle_{x}$
measurement outcome; orange: the $|1\rangle_{z}|-1\rangle_{x}$ measurement
outcome; green: the $|-1\rangle_{z}|1\rangle_{x}$ measurement outcome;
and blue the $|-1\rangle_{z}|-1\rangle_{x}$ measurement outcome.
Initial state: $|\Psi^{-}\rangle=\frac{1}{\sqrt{2}}(|1\rangle_{z}|-1\rangle_{z}-|-1\rangle_{z}|1\rangle_{z})$,
measurements strengths $a_{1}=a_{2}=1$, time-step 0.0001. \label{fig:z_x_prob_amps}}

\end{figure}

The petal pattern appears as a result of a multi-stage collapse cascade
depicted in Figure \ref{fig:z_x_prob_amps}. The system is initiated
with non-zero probability amplitudes for all four possible measurement
outcomes but this reduces to three, then two and finally just one
significant contribution. The system evolves towards one of the superposition
circles depicted in Figure \ref{fig:petal_superpositions} and eventually
to the tip of the corresponding petal. For the probability amplitudes
depicted in Figure \ref{fig:z_x_prob_amps}, the trajectory approaches
the blue circle in Figure \ref{fig:petal_superpositions} defined
by non-zero probability amplitudes for the $|1\rangle_{z}|1\rangle_{x}$
and $|-1\rangle_{z}|1\rangle_{x}$ measurement outcomes. The system
then travels along the circle and collapses at one of these measurement
outcomes: in Figure \ref{fig:z_x_prob_amps} it is the probability
amplitude for the $|1\rangle_{z}|1\rangle_{x}$ measurement outcome
that approaches unity. The regions outside the petals remain unexplored
since the system begins at the centre, $\langle S_{x,2}\rangle=\langle S_{z,1}\rangle=0$,
and cannot pass across a superposition circle.

\begin{figure}
\begin{centering}
\includegraphics[width=1\columnwidth]{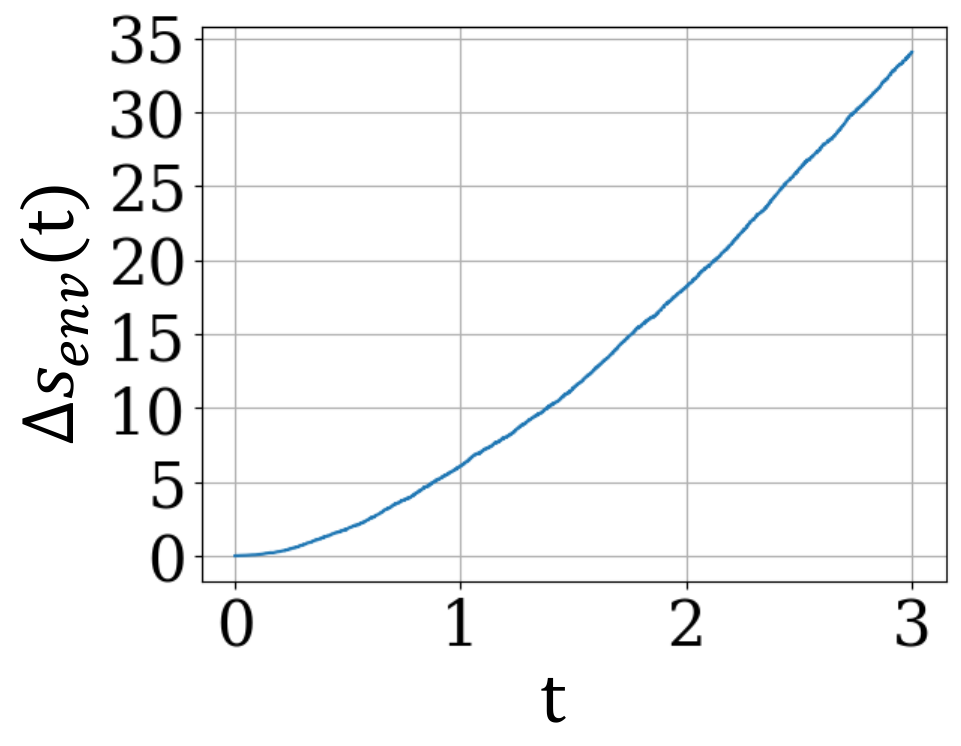}
\par\end{centering}
\caption{The mean environmental stochastic entropy production for Case 2:
two spin 1/2 particles undergoing a $z$-spin measurement on particle
1 and an $x$-spin measurement on particle 2. Initial state: $|\Psi^{-}\rangle=\frac{1}{\sqrt{2}}(|1\rangle_{z}|-1\rangle_{z}-|-1\rangle_{z}|1\rangle_{z})$,
314 trajectories, measurement strengths $a_{1}=a_{2}=1$, time-step
0.0001. \label{fig:z_x_ent_prod}}
\end{figure}

The mean environmental stochastic entropy production associated with
Case 2 is depicted in Figure \ref{fig:z_x_ent_prod}. After an initially
slow rate of entropy production, the mean rate rises to around three
to four times that of a single particle undergoing a $z$-spin measurement
and nearly double that of Case 1. The rate of entropy production for
Case 2 is higher than for Case 1 possibly because there are four rather
than two potential measurement outcomes the system could adopt. Furthermore,
in Case 1, the system already starts on a superposition circle, while
in Case 2 the system requires time to reach a superposition circle
and complete the first stage of the measurement process. This could
explain the initially slow rate of production exhibited by Case 2.

\section{Measurement of total $Z$-spin of two spin 1/2 particles \label{sec:global_mes}}

\subsection{Dynamics}

We now perform measurements of the $z$ component of the total spin
of the two spin 1/2 system: $S_{z}=\frac{1}{2}\sigma_{z,1}\otimes\mathbb{I}+\mathbb{I}\otimes\frac{1}{2}\sigma_{z,2}$
where subscripts $1$ and $2$ label the two particles. The Lindblad
operator $L_{z}=aS_{z}$, where $a$ is a coupling strength, is used
in Eq. (\ref{eq:SME-1}) to obtain SDEs for the fifteen variables
parametrising the density matrix. The full set can be found in Appendix
\ref{sec:Appendix-G}. It is convenient, however, to consider the
evolution in terms of fewer variables. Using $\langle X\rangle={\rm Tr}(X\rho),$
we obtain
\begin{align}
\langle S_{z}\rangle & =\frac{s_{3}+s_{12}}{2},\qquad\langle S_{x}\rangle=\frac{s_{4}+s_{1}}{2}\nonumber \\
\langle S_{y}\rangle & =\frac{s_{2}+s_{8}}{2},\quad\langle S^{2}\rangle=\frac{1}{2}(3+s_{5}+s_{10}+s_{15})\nonumber \\
\langle S_{z,1}\rangle & =\frac{1}{2}s_{12},\qquad\langle S_{z,2}\rangle=\frac{1}{2}s_{3},\label{eq:spin_comp_variables}
\end{align}
where $\langle S_{x}\rangle$ and $\langle S_{y}\rangle$ are defined
in a similar manner to $\langle S_{z}\rangle$,  $S^{2}=S_{x}^{2}+S_{y}^{2}+S_{z}^{2}$,
and $\langle S_{z,1}\rangle$ and $\langle S_{z,2}\rangle$ are the
$z$-spins of particle 1 and particle 2, respectively.

 By noting that $d\langle X\rangle={\rm Tr}(Xd\rho)$, we can write
the SDE for $\langle S^{2}\rangle$ as follows:

\begin{align}
d\langle S^{2}\rangle & =\frac{1}{2}a(s_{3}+s_{12})(1-s_{5}-s_{10}-s_{15})dW_{z}.\label{eq:25}
\end{align}

Note that in three of the five Cases that follow, we restrict the
dynamics by starting the system in one of the triplet eigenstates
of $S^{2}$ with $\langle S^{2}\rangle=2$, or a superposition of
them. For a general superposition of triplet eigenstates (with real
amplitudes) we can compute the $s_{i}$ using $s_{i}={\rm Tr}(\rho_{{\rm trip}}\Sigma_{i})$,
where the state is represented by $\rho_{{\rm trip}}$, and show that
$s_{5}+s_{10}+s_{15}=1$. Thus the second bracket in Eq. (\ref{eq:25})
is zero and $d\langle S^{2}\rangle=0$ throughout, implying that $\langle S^{2}\rangle$
remains constant and the system can never reach the singlet state
at $\langle S^{2}\rangle=0$. We consider real amplitudes only since
for the initial states we explore in this section, the variables $s_{2}=s_{6}=s_{8}=s_{9}=s_{11}=s_{14}=0$,
and remain so for all of time since $ds_{2}=ds_{6}=ds_{8}=ds_{9}=ds_{11}=ds_{14}=0$.
Therefore only the real elements in the general expression for $\rho$,
Eq. (\ref{eq:ent_spin_density_matrix}), are non-zero. When starting
in an initial state that is a triplet eigenstate of $S^{2}$, the
system of two spin 1/2 particles is therefore equivalent to a single
spin 1 system, and the three possible eigenstates of $S_{z}$ are
$|1\rangle_{z}|1\rangle_{z}$ corresponding to $\langle S_{z}\rangle=1$;
$\frac{1}{\sqrt{2}}(|1\rangle_{z}|-1\rangle_{z}+|-1\rangle_{z}|1\rangle_{z})$
corresponding to $\langle S_{z}\rangle=0$; and $|-1\rangle_{z}|-1\rangle_{z}$
corresponding to $\langle S_{z}\rangle=-1$.

\subsection{Environmental stochastic entropy production}

The environmental stochastic entropy production is again calculated
using an appropriate reduced diffusion matrix. We note that there
is a closed set of three SDEs associated with the protocol of $S_{z}$
measurement:

\begin{align}
ds_{12} & =a(-s_{12}^{2}-s_{3}s_{12}+s_{15}+1)dW_{z}=-a\kappa dW_{z}\nonumber \\
ds_{3} & =a(-s_{3}^{2}-s_{3}s_{12}+s_{15}+1)dW_{z}=-a\nu dW_{z}\nonumber \\
ds_{15} & =a(-s_{3}s_{15}+s_{3}-s_{12}s_{15}+s_{12})dW_{z}=-a\gamma dW_{z},
\end{align}
where compact notation in terms of $\kappa,$ $\nu$ and $\gamma$
is introduced. The reduced diffusion matrix is given by

\begin{align*}
\boldsymbol{D}_{S_{z}} & =\frac{1}{2}a^{2}\left(\begin{array}{ccc}
\nu^{2} & \nu\kappa & \nu\gamma\\
\nu\kappa & \kappa^{2} & \nu\gamma\\
\nu\gamma & \kappa\gamma & \gamma^{2}
\end{array}\right),
\end{align*}
 and since this matrix is singular, we follow the procedure outlined
in Section \ref{subsec:Singular-Diffusion-Matrices}. The diffusion
matrix has only one non-zero eigenvalue, and so we require a single
dynamical variable. For simplicity, we chose $s_{3}$ to be the dynamical
variable with $s_{12}$ and $s_{15}$ acting as spectator variables.
The scalar diffusion coefficient is then $D_{S_{z}}=\frac{1}{2}a^{2}\nu^{2}$.
The environmental stochastic entropy production is found using the
correction terms for the derivatives in Eq. (\ref{eq:corr_terms})
and Eq. (\ref{eq:1d_stoch_ent-1}) and may be expressed as follows:

\begin{widetext}

\begin{align}
d\Delta s_{{\rm env}} & =a^{2}\left[-\frac{1}{\nu}\left(-(4s_{3}+2s_{12})\gamma+(4s_{3}^{2}+4s_{3}s_{12}-2s_{15})\kappa\right)-(6s_{3}^{2}+8s_{3}s_{12}+2s_{12}^{2}-4s_{15}-2)\right.\nonumber \\
 & \left.+\frac{1}{\nu^{2}}\left(16s_{3}^{3}+32s_{3}^{2}s_{12}+16s_{3}s_{12}^{2}-32s_{3}s_{15}-16s_{3}-16s_{12}s_{15}\right)^{2}\right]dt\nonumber \\
 & +\frac{1}{\nu}a\left(4s_{3}^{3}+8s_{3}^{2}s_{12}+4s_{3}s_{12}^{2}-8s_{3}s_{15}-4s_{3}-4s_{12}s_{15}\right)dW_{z}.\label{eq:ent_prod_2_spin_half_Sz}
\end{align}

\end{widetext}Note that when the system is close to an eigenstate,
the expression for $d\Delta s_{{\rm env}}$ is vulnerable to exhibiting
singularities due to numerical inaccuracies. The longer the duration
of measurement, the more likely this will occur. Trajectories that
gave rise to singularities were not considered in the calculation
of mean environmental entropy production. 

\subsection{Results for various cases}

\begin{figure}
\begin{centering}
\includegraphics[width=1\columnwidth]{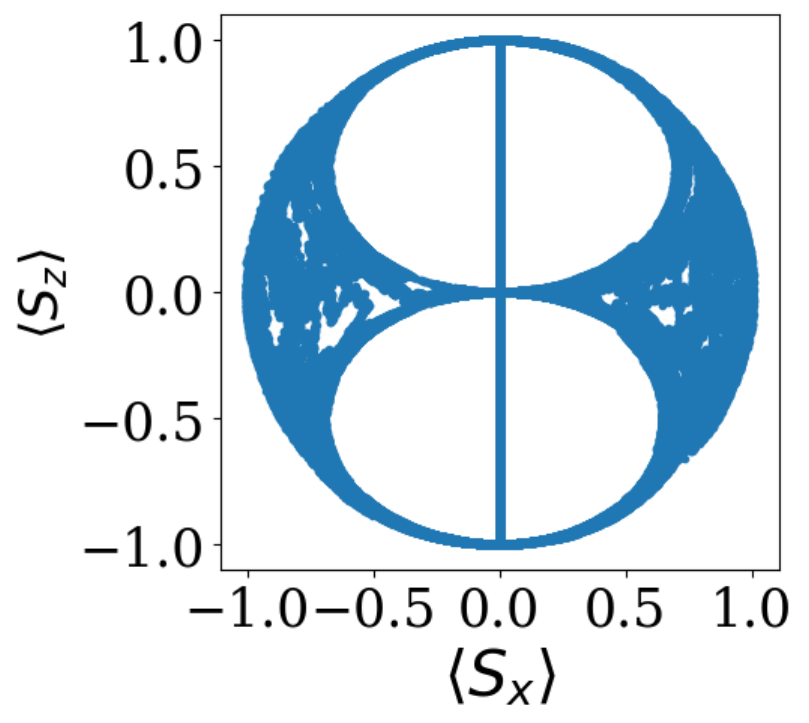}
\par\end{centering}
\caption{Stochastic evolution of the system of two spin 1/2 particles undergoing
measurement of total $S_{z}$. The 50 trajectories have been illustrated
in terms of the spin components $\langle S_{x}\rangle,\langle S_{z}\rangle$.
Each trajectory starts in one of the three triplet eigenstate of $S_{x}$
represented as the purple and grey crosses in Figure \ref{fig:sz_sx_superpositions}
and terminates in the vicinity of one of the triplet eigenstates of
$S_{z}$ at the top, bottom and centre of the plot. Measurement strength
$a=1$, measurement duration 6 time units with time-step 0.0001.\label{fig:2_spin_half_ex_traj}
}

\end{figure}

\begin{figure}
\begin{centering}
\includegraphics[width=1\columnwidth]{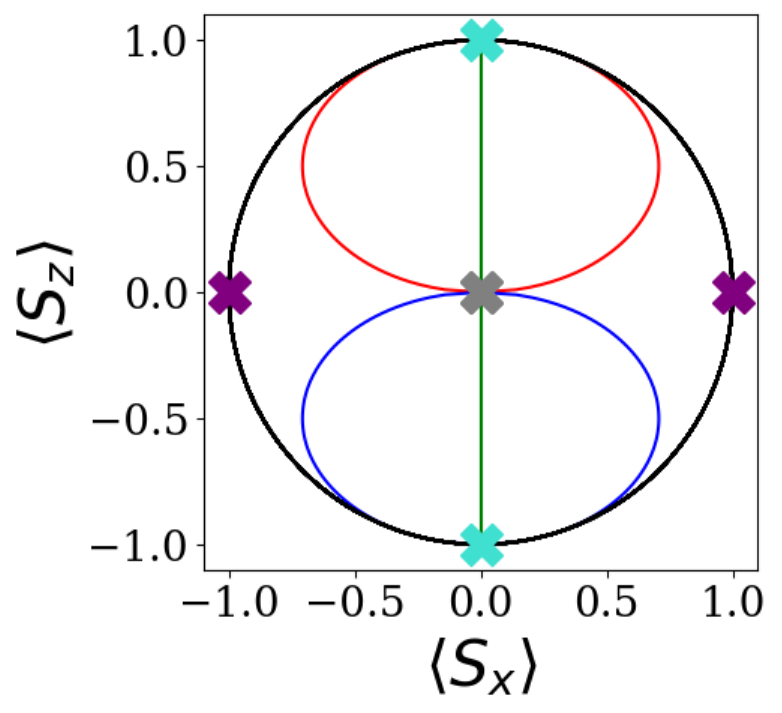}
\par\end{centering}
\caption{The curves represent superpositions (with real amplitudes) between
different eigenstates of $S_{z}$. The upper and lower ellipses describe
a superposition between the $|1\rangle_{z}|1\rangle_{z}$ and $\frac{1}{\sqrt{2}}(|1\rangle_{z}|-1\rangle_{z}+|-1\rangle_{z}|1\rangle_{z})$
eigenstates (red), and the $|-1\rangle_{z}|-1\rangle_{z}$ and $\frac{1}{\sqrt{2}}(|1\rangle_{z}|-1\rangle_{z}+|-1\rangle_{z}|1\rangle_{z})$
eigenstates (blue), respectively. The vertical green line represents
a superposition between the $|-1\rangle_{z}|-1\rangle_{z}$ and $|1\rangle_{z}|1\rangle_{z}$
eigenstates \citep{walls2024}. The purple crosses represent the $|1\rangle_{x}|1\rangle_{x}$
and $|-1\rangle_{x}|-1\rangle_{x}$ eigenstates of $S_{x}$ and the
turquoise crosses the $|1\rangle_{z}|1\rangle_{z}$ and $|-1\rangle_{z}|-1\rangle_{z}$
eigenstates of $S_{z}$. The grey cross represents both the $\frac{1}{\sqrt{2}}(|1\rangle_{x}|-1\rangle_{x}+|-1\rangle_{x}|1\rangle_{x})$
and $\frac{1}{\sqrt{2}}(|1\rangle_{z}|-1\rangle_{z}+|-1\rangle_{z}|1\rangle_{z})$
eigenstates, noting they are orthogonal. \label{fig:sz_sx_superpositions}}

\end{figure}

Stochastic trajectories are generated as the two spin 1/2 system undergoes
a measurement of $S_{z}$ and Figure \ref{fig:2_spin_half_ex_traj}
depicts example trajectories in the $\langle S_{x}\rangle,\langle S_{z}\rangle$
co-ordinate space. We first initiate the system in one of the three
triplet eigenstates of $S_{x}$: either the $|1\rangle_{x}|1\rangle_{x}$,
$|-1\rangle_{x}|-1\rangle_{x}$ or the $\frac{1}{\sqrt{2}}(|1\rangle_{x}|-1\rangle_{x}+|-1\rangle_{x}|1\rangle_{x})$
eigenstate (represented as the two purple and one grey crosses in
Figure \ref{fig:sz_sx_superpositions}) or any superposition of the
three triplet eigenstates of $S_{z}$, such that the system remains
in the triplet state space with $\langle S^{2}\rangle=2$. 

The $|1\rangle_{z}|1\rangle_{z}$ and $|-1\rangle_{z}|-1\rangle_{z}$
eigenstates lie at the turquoise crosses at the top and bottom of
Figure \ref{fig:sz_sx_superpositions}. The $\frac{1}{\sqrt{2}}(|1\rangle_{z}|-1\rangle_{z}+|-1\rangle_{z}|1\rangle_{z})$
and $\frac{1}{\sqrt{2}}(|1\rangle_{x}|-1\rangle_{x}+|-1\rangle_{x}|1\rangle_{x})$
eigenstates lie at the centre of the plot (the grey cross in Figure
\ref{fig:sz_sx_superpositions}) though they are orthogonal. The avoidance
of the figure-of-eight-like region in Figure \ref{fig:2_spin_half_ex_traj}
can be understood using a measurement collapse cascade model \citep{walls2024}
and the loci of superpositions (with real amplitudes) between pairs
of eigenstates of $S_{z}$. The red and blue ellipses in Figure \ref{fig:sz_sx_superpositions}
describe a superposition between the $|1\rangle_{z}|1\rangle_{z}$
and $\frac{1}{\sqrt{2}}(|1\rangle_{z}|-1\rangle_{z}+|-1\rangle_{z}|1\rangle_{z})$
eigenstates, and the $|-1\rangle_{z}|-1\rangle_{z}$ and $\frac{1}{\sqrt{2}}(|1\rangle_{z}|-1\rangle_{z}+|-1\rangle_{z}|1\rangle_{z})$
eigenstates, respectively. The green line represents a superposition
between the $|-1\rangle_{z}|-1\rangle_{z}$ and $|1\rangle_{z}|1\rangle_{z}$
eigenstates. The black outer circle represents a further evolution
pathway for collapse to the $|-1\rangle_{z}|-1\rangle_{z}$ and $|1\rangle_{z}|1\rangle_{z}$
eigenstates.

The empty regions result from a two-step measurement collapse cascade.
Starting from the $|-1\rangle_{x}|-1\rangle_{x}$ eigenstate, for
example, the system has non-zero probability amplitudes for collapse
to any of the three eigenstates of $S_{z}$. The decrease of one of
these probability amplitudes to zero is the first step in the collapse
process, such that the system arrives at one of the superposition
loci, either the red or the blue ellipse. The system is not able to
recover a non-zero probability amplitude for the third eigenstate,
and it then moves along the ellipse towards one of the two eigenstates
to complete the collapse. The empty regions cannot be reached from
a starting point at one of the eigenstates of $S_{x}$ since this
would involve crossing a superposition locus. Further discussion of
this pattern formation can be found in Walls and Ford \citep{walls2024}.

Note the similarities between the avoidance of regions in Figure \ref{fig:2_spin_half_ex_traj}
for the system undergoing measurements of $S_{z}$, and the pattern
formation in Figure \ref{fig:z_x_example_traj} for simultaneous measurement
of $S_{x,2}$ and $S_{z,1}$.

We now consider the environmental stochastic entropy production associated
with measurement of total $S_{z}$ of the two spin 1/2 system. In
particular, we compare the mean rates of asymptotic environmental
stochastic entropy production when the system is prepared in different
initial states and collapses to various eigenstates of $S_{z}$.

\subsection*{Case A. }

\begin{figure}
\begin{centering}
\includegraphics[width=1\columnwidth]{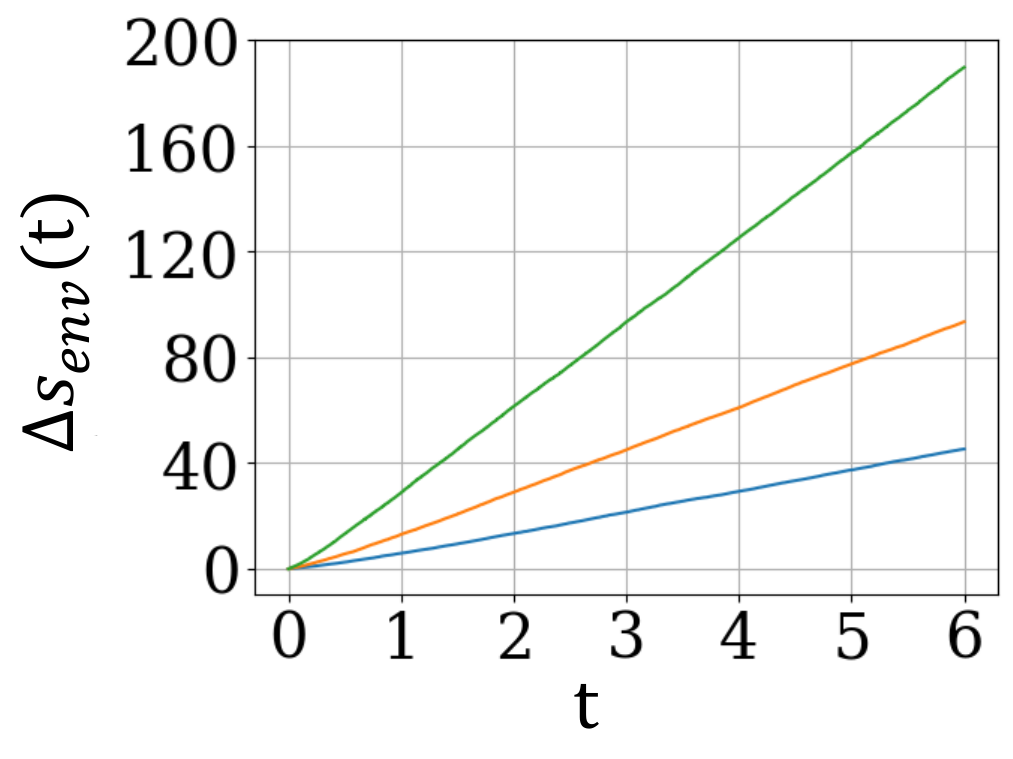}
\par\end{centering}
\caption{The mean environmental stochastic entropy production for Case A: the
two spin 1/2 system undergoing $S_{z}$ measurement starting from
the $\frac{1}{\sqrt{2}}(|1\rangle_{x}|-1\rangle_{x}+|-1\rangle_{x}|1\rangle_{x})$
eigenstate and approaching the $|1\rangle_{z}|1\rangle_{z}$ eigenstate.
Measurement strengths: in blue: $a=\frac{1}{\sqrt{2}}$, in orange:
$a=1$, and in green: $a=\sqrt{2}$. 600 trajectories, time-step 0.0001.
\label{fig:2_spin_half_entropy_start_Sx_0}}

\end{figure}

We first consider the mean environmental stochastic entropy production
when the system starts in the $\frac{1}{\sqrt{2}}(|1\rangle_{x}|-1\rangle_{x}+|-1\rangle_{x}|1\rangle_{x})$
eigenstate. Note that from this initial state, there is zero probability
of collapsing to the $\frac{1}{\sqrt{2}}(|1\rangle_{z}|-1\rangle_{z}+|-1\rangle_{z}|1\rangle_{z})$
eigenstate of $S_{z}$ and but equal probabilities of collapsing to
$|1\rangle_{z}|1\rangle_{z}$ and $|-1\rangle_{z}|-1\rangle_{z}$.
The system already lies on the superposition locus defined by the
vertical green line in Figure \ref{fig:sz_sx_superpositions} and
therefore does not need to undergo the first stage of the measurement
collapse cascade. Figure \ref{fig:2_spin_half_entropy_start_Sx_0}
illustrates the mean environmental stochastic entropy production when
approaching the $|1\rangle_{z}|1\rangle_{z}$ eigenstate, for different
measurement strengths. A similar plot would be produced for approach
to the $|-1\rangle_{z}|-1\rangle_{z}$ eigenstate.The mean rate of
environmental stochastic entropy production is roughly proportional
to the square of the measurement strength $a$.

Starting from this initial state means that $s_{15}=1$ throughout
and, together with $s_{3}=s_{12}$, the environmental stochastic entropy
production in Eq. (\ref{eq:ent_prod_2_spin_half_Sz}) simplifies to
$d\Delta s_{{\rm env}}=8a^{2}(s_{12}^{2}+1)dt+8as_{12}dW_{z}$. We
would thus expect the asymptotic mean rate of stochastic environmental
entropy production to be 16 for a measurement strength $a=1$, which
is consistent with the orange curve in Figure \ref{fig:2_spin_half_entropy_start_Sx_0}.
Note that this is double that of the single particle $z$-spin measurements
performed on the two spin 1/2 system (Case 1), depicted in Figure
\ref{fig:z_z_mes_ent}.

\subsection*{Case B.}

\begin{figure}
\begin{centering}
\includegraphics[width=1\columnwidth]{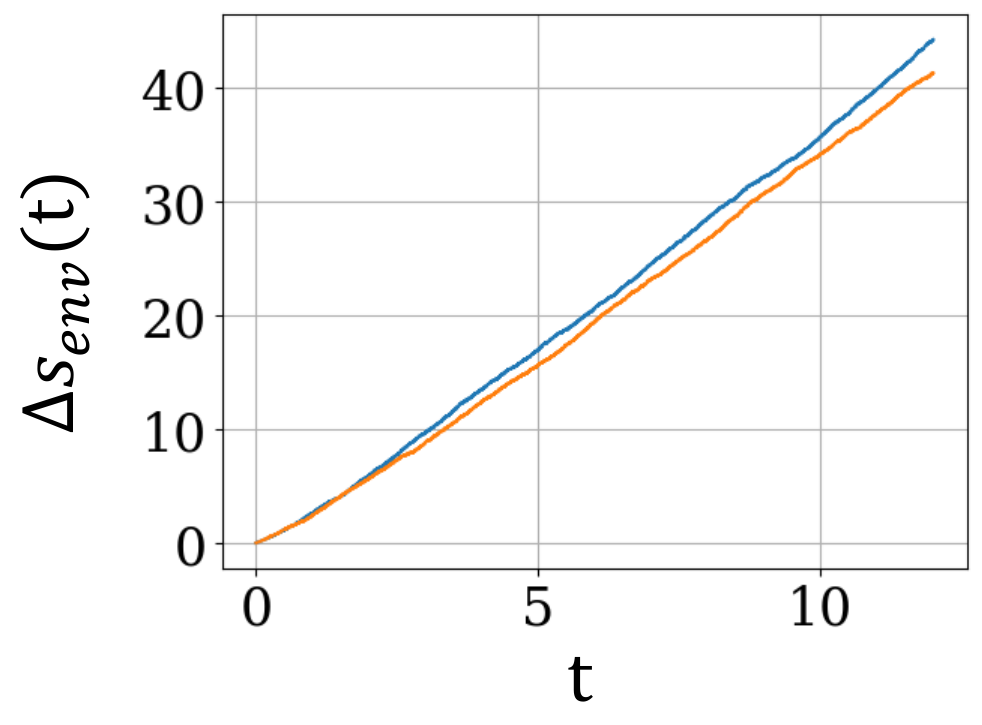}
\par\end{centering}
\caption{The mean environmental stochastic entropy production for Case B: the
system undergoing a measurement of $S_{z}$ starting in $\frac{1}{\sqrt{2}}|1\rangle_{z}|1\rangle_{z}+\frac{1}{\sqrt{2}}(|1\rangle_{z}|-1\rangle_{z}+|-1\rangle_{z}|1\rangle_{z})$:
an equal superposition (with real and positive amplitudes) of the
$|1\rangle_{z}|1\rangle_{z}$ and $\frac{1}{\sqrt{2}}(|1\rangle_{z}|-1\rangle_{z}+|-1\rangle_{z}|1\rangle_{z})$
eigenstates. The production for collapse to the $|1\rangle_{z}|1\rangle_{z}$
eigenstate is shown in blue and for collapse to the $\frac{1}{\sqrt{2}}(|1\rangle_{z}|-1\rangle_{z}+|-1\rangle_{z}|1\rangle_{z})$
eigenstate in orange. 370 trajectories, $a=1$, measurement duration
12, time-step 0.0001. \label{fig:2_spin_half_ent_prod_eq_sup_sz_tripzero_sz_1}}

\end{figure}

We go on to consider a situation where the system takes an initial
state $\frac{1}{\sqrt{2}}|1\rangle_{z}|1\rangle_{z}+\frac{1}{\sqrt{2}}(|1\rangle_{z}|-1\rangle_{z}+|-1\rangle_{z}|1\rangle_{z})$,
hence with equal probabilities of collapsing at either the $|1\rangle_{z}|1\rangle_{z}$
or the $\frac{1}{\sqrt{2}}(|1\rangle_{z}|-1\rangle_{z}+|-1\rangle_{z}|1\rangle_{z})$
eigenstates of $S_{z}$. This is located at $\langle S_{x}\rangle=\frac{1}{\sqrt{2}},\langle S_{z}\rangle=\frac{1}{2}$
on the red ellipse in Figure \ref{fig:sz_sx_superpositions}, and
therefore the system proceeds without need for the first stage of
the measurement collapse cascade. The mean environmental stochastic
entropy production associated with collapse to the $|1\rangle_{z}|1\rangle_{z}$
and $\frac{1}{\sqrt{2}}(|1\rangle_{z}|-1\rangle_{z}+|-1\rangle_{z}|1\rangle_{z})$
eigenstates is depicted in Figure \ref{fig:2_spin_half_ent_prod_eq_sup_sz_tripzero_sz_1},
illustrated as the blue and orange lines respectively. The mean asymptotic
rate of stochastic entropy production for approach to both eigenstates
is around  four times smaller than that associated with collapse
at the $|1\rangle_{z}|1\rangle_{z}$ and $|-1\rangle_{z}|-1\rangle_{z}$
eigenstates in Case A starting from $\frac{1}{\sqrt{2}}(|1\rangle_{x}|-1\rangle_{x}+|-1\rangle_{x}|1\rangle_{x})$
for $a=1$, shown in Figure \ref{fig:2_spin_half_entropy_start_Sx_0}.

We speculate that collapse to the $|1\rangle_{z}|1\rangle_{z}$ and
$|-1\rangle_{z}|-1\rangle_{z}$ eigenstates in Case A (Figure \ref{fig:2_spin_half_entropy_start_Sx_0})
occurs more quickly than collapse in Case B because the system already
begins with $s_{15}=1$, the value required for the system to take
either of the $|1\rangle_{z}|1\rangle_{z}$ and $|-1\rangle_{z}|-1\rangle_{z}$
eigenstates, and greater speed is associated with a higher rate of
mean asymptotic entropy production. In contrast, the system in Case
B (Figure \ref{fig:2_spin_half_ent_prod_eq_sup_sz_tripzero_sz_1})
is initiated with $s_{15}=0$ while $s_{15}=1$ is required for the
system to take the $|1\rangle_{z}|1\rangle_{z}$ eigenstate and $s_{15}=-1$
for the $\frac{1}{\sqrt{2}}(|1\rangle_{z}|-1\rangle_{z}+|-1\rangle_{z}|1\rangle_{z})$
eigenstate. The system therefore requires more time to reach the desired
value of $s_{15}$ in Case B, and the asymptotic mean rate of environmental
stochastic entropy production could be slower than Case A because
of this.

\subsection*{Case C.}

\begin{figure}
\begin{centering}
\includegraphics[width=1\columnwidth]{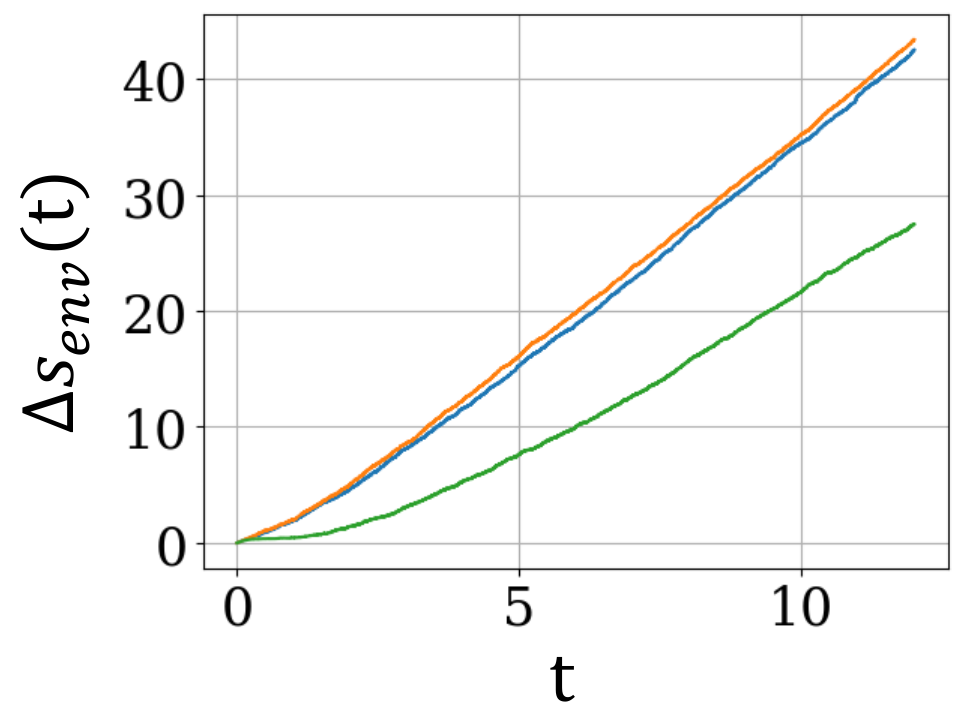}
\par\end{centering}
\caption{The mean environmental stochastic entropy production for Case C: approach
towards the $|1\rangle_{z}|1\rangle_{z}$ eigenstate (blue), $|-1\rangle_{z}|-1\rangle_{z}$
eigenstate (orange) and $\frac{1}{\sqrt{2}}(|1\rangle_{z}|-1\rangle_{z}+|-1\rangle_{z}|1\rangle_{z})$
eigenstate (green) during an $S_{z}$ measurement starting from the
$|1\rangle_{x}|1\rangle_{x}$ eigenstate. Measurement strength $a=1$,
measurement duration 12. 370 trajectories, time-step 0.0001. \label{fig:The-average-stochastic2_spin_half_entropy_start_Sx_1}}

\end{figure}

The next initial state we consider is the $|1\rangle_{x}|1\rangle_{x}$
eigenstate from which the system can reach either the $|1\rangle_{z}|1\rangle_{z},|-1\rangle_{z}|-1\rangle_{z}$
or $\frac{1}{\sqrt{2}}(|1\rangle_{z}|-1\rangle_{z}+|-1\rangle_{z}|1\rangle_{z})$
eigenstates. The system must now undergo the first stage of the measurement
collapse cascade since it does not start on one of the superposition
loci depicted in Figure \ref{fig:sz_sx_superpositions}. Figure \ref{fig:The-average-stochastic2_spin_half_entropy_start_Sx_1}
illustrates the mean environmental stochastic entropy production associated
with approach to the $|1\rangle_{z}|1\rangle_{z}$ eigenstate (blue),
the $|-1\rangle_{z}|-1\rangle_{z}$ eigenstate (orange) and the $\frac{1}{\sqrt{2}}(|1\rangle_{z}|-1\rangle_{z}+|-1\rangle_{z}|1\rangle_{z})$
eigenstate (green), starting from the $|1\rangle_{x}|1\rangle_{x}$
eigenstate.

\begin{figure}
\begin{centering}
\includegraphics[width=1\columnwidth]{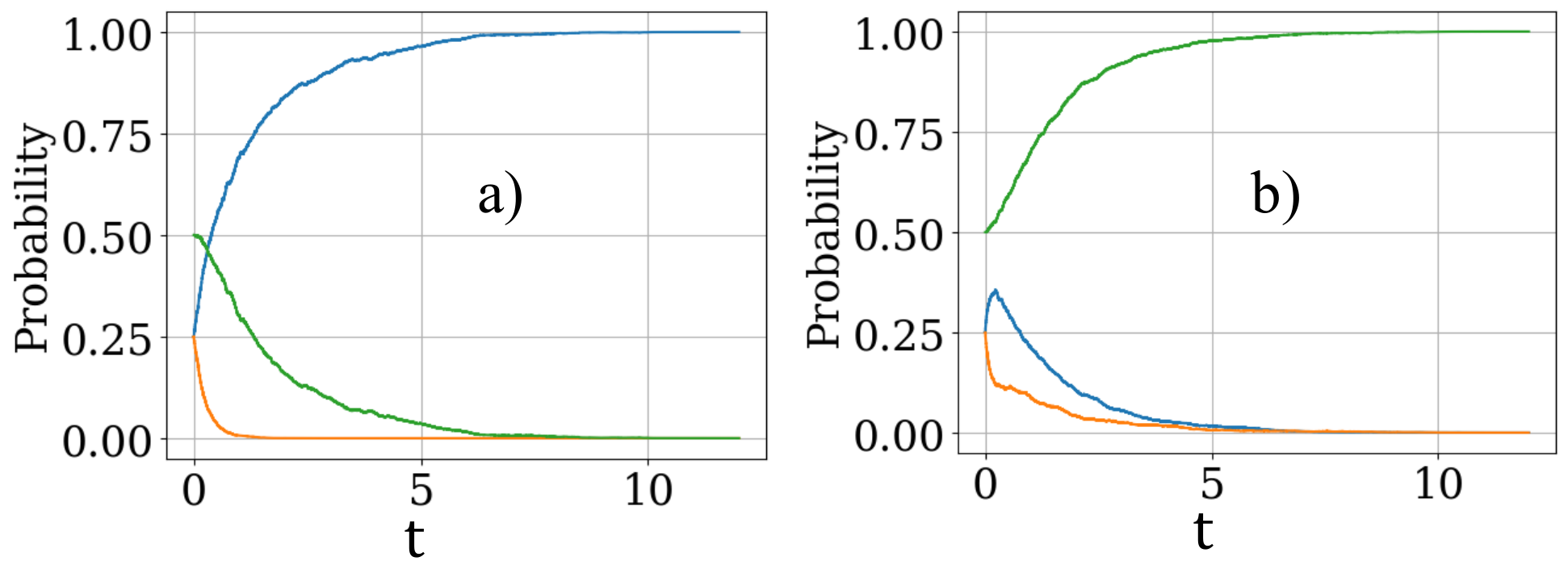}
\par\end{centering}
\caption{The mean variation of the probability amplitudes of the triplet eigenstates
of $S_{z}$, conditioned on collapse at a specific eigenstate. Their
variation is shown over the duration of a measurement for Case C,
$a=1$ and 370 trajectories.  Blue represents the probability amplitude
for the $|-1\rangle_{z}|-1\rangle_{z}$ eigenstate, green the $\frac{1}{\sqrt{2}}(|1\rangle_{z}|-1\rangle_{z}+|-1\rangle_{z}|1\rangle_{z})$
eigenstate and orange the $|1\rangle_{z}|1\rangle_{z}$ eigenstate.
a) The system is conditioned on collapse at the $|-1\rangle_{z}|-1\rangle_{z}$
eigenstate.  b) The system is conditioned on collapse at the $\frac{1}{\sqrt{2}}(|1\rangle_{z}|-1\rangle_{z}+|-1\rangle_{z}|1\rangle_{z})$
eigenstate and the $|1\rangle_{z}|1\rangle_{z}$ eigenstate being
first probability amplitude to fall to zero, 445 trajectories. \label{fig:prob_amps_av}}

\end{figure}

There appears to be a lower mean rate of asymptotic environmental
entropy production associated with collapse to the $\frac{1}{\sqrt{2}}(|1\rangle_{z}|-1\rangle_{z}+|-1\rangle_{z}|1\rangle_{z})$
eigenstate compared with the $|1\rangle_{z}|1\rangle_{z}$ and $|-1\rangle_{z}|-1\rangle_{z}$
eigenstates. To investigate further, we consider the mean variation
of the probability amplitudes of the triplet eigenstates of $S_{z}$,
conditioned on collapse at a specific eigenstate. Figure \ref{fig:prob_amps_av}a
illustrates the mean variation of these amplitudes conditioned on
system collapse at the $|-1\rangle_{z}|-1\rangle_{z}$ eigenstate,
whilst in Figure \ref{fig:prob_amps_av}b the trajectories are conditioned
on the system collapsing at the $\frac{1}{\sqrt{2}}(|1\rangle_{z}|-1\rangle_{z}+|-1\rangle_{z}|1\rangle_{z})$
eigenstate and the $|1\rangle_{z}|1\rangle_{z}$ eigenstate being
the first probability amplitude to go to zero. The blue line represents
the probability amplitude of the $|-1\rangle_{z}|-1\rangle_{z}$ eigenstate,
the green line the $\frac{1}{\sqrt{2}}(|1\rangle_{z}|-1\rangle_{z}+|-1\rangle_{z}|1\rangle_{z})$
eigenstate and the orange line the $|1\rangle_{z}|1\rangle_{z}$ eigenstate.

In both Figures \ref{fig:prob_amps_av}a and \ref{fig:prob_amps_av}b,
the $|1\rangle_{z}|1\rangle_{z}$ eigenstate is the first probability
amplitude to fall to zero, thus we can compare how long it takes
for this first stage of the collapse cascade to be completed. In Figure
\ref{fig:prob_amps_av}a the probability amplitude of the $|1\rangle_{z}|1\rangle_{z}$
eigenstate reaches zero at around 1 time-unit, whereas in Figure \ref{fig:prob_amps_av}b
it only reaches zero at around 6 time-units.  Moreover, in Figure
\ref{fig:prob_amps_av}b, the probability amplitude of the $|-1\rangle_{z}|-1\rangle_{z}$
eigenstate initially rises before falling towards zero, revealing
perhaps that the system experiences a greater attraction towards the
$|\pm1\rangle_{z}|\pm1\rangle_{z}$ eigenstates of $S_{z}$ and more
difficulty in collapsing to the $\frac{1}{\sqrt{2}}(|1\rangle_{z}|-1\rangle_{z}+|-1\rangle_{z}|1\rangle_{z})$
eigenstate. Perhaps this greater competition arises because the $\frac{1}{\sqrt{2}}(|1\rangle_{z}|-1\rangle_{z}+|-1\rangle_{z}|1\rangle_{z})$
eigenstate lies in the middle of the ladder of eigenstates of $S_{z}$
and at the centre of the circle in the $\langle S_{x}\rangle,\langle S_{z}\rangle$
coordinate space in Figure \ref{fig:2_spin_half_ex_traj}. In contrast,
the $|\pm1\rangle_{z}|\pm1\rangle_{z}$ eigenstates lie at the termini
of the ladder and at the top or bottom of the circle in the $\langle S_{x}\rangle,\langle S_{z}\rangle$
coordinate space: therefore they only face competition from the $\frac{1}{\sqrt{2}}(|1\rangle_{z}|-1\rangle_{z}+|-1\rangle_{z}|1\rangle_{z})$
eigenstate and not from each other. 

The asymptotic mean rates of environmental stochastic entropy production
for the approach to the $\frac{1}{\sqrt{2}}(|1\rangle_{z}|-1\rangle_{z}+|-1\rangle_{z}|1\rangle_{z})$
eigenstate differ when the system starts in the $|1\rangle_{x}|1\rangle_{x}$
eigenstate (Case C) or in the $\frac{1}{\sqrt{2}}(|1\rangle_{z}|1\rangle_{z}+\frac{1}{\sqrt{2}}(|1\rangle_{z}|-1\rangle_{z}+|-1\rangle_{z}|1\rangle_{z})$
state (Case B). This is not the case for the $|\pm1\rangle_{z}|\pm1\rangle_{z}$
eigenstates which exhibit the same asymptotic mean rates in Case B
and Case C. This could again arise because the $\frac{1}{\sqrt{2}}(|1\rangle_{z}|-1\rangle_{z}+|-1\rangle_{z}|1\rangle_{z})$
eigenstate lies between the other two.  In Case B the $\frac{1}{\sqrt{2}}(|1\rangle_{z}|-1\rangle_{z}+|-1\rangle_{z}|1\rangle_{z})$
eigenstate only has to compete in the collapse with the $|1\rangle_{z}|1\rangle_{z}$
eigenstate, but in Case C the $\frac{1}{\sqrt{2}}(|1\rangle_{z}|-1\rangle_{z}+|-1\rangle_{z}|1\rangle_{z})$
eigenstate competes for selection with both $|1\rangle_{z}|1\rangle_{z}$
and $|-1\rangle_{z}|-1\rangle_{z}$ and thus it is more affected by
a non-zero probability amplitude of the $|-1\rangle_{z}|-1\rangle_{z}$
eigenstate in Case C.

\subsection*{Case D.}

\begin{figure}
\begin{centering}
\includegraphics[width=1\columnwidth]{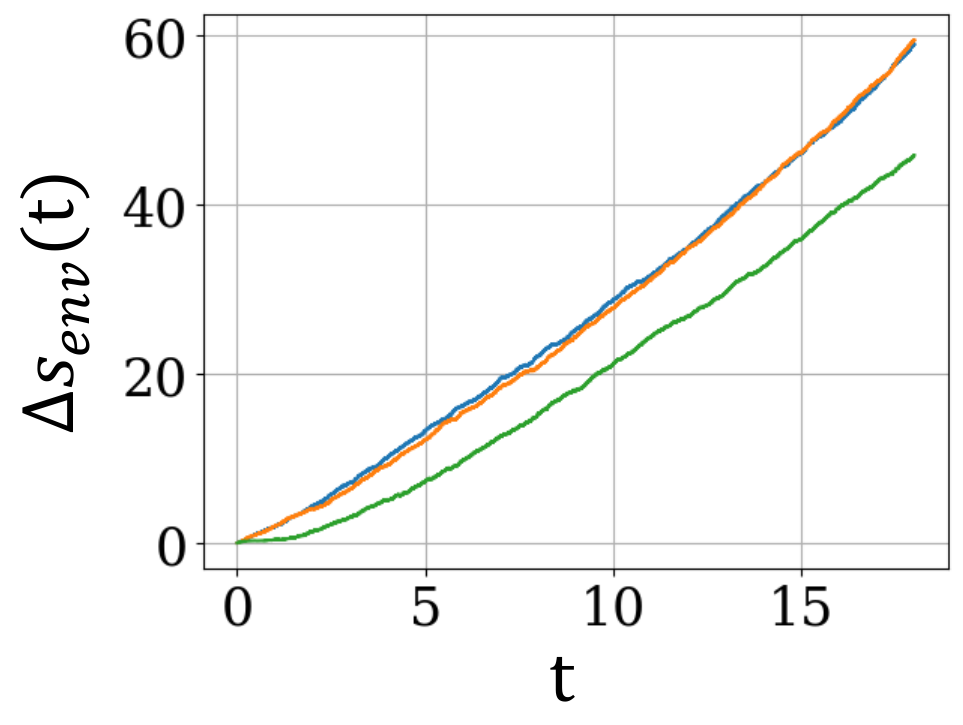}
\par\end{centering}
\caption{The mean environmental stochastic entropy production for Case D: the
system starts in a mixed state $\rho_{{\rm in}}=\frac{1}{4}(|1\rangle_{z}|1\rangle_{z}\langle1|_{z}\langle1|_{z}+|-1\rangle_{z}|-1\rangle_{z}\langle-1|_{z}\langle-1|_{z}+|1\rangle_{z}|-1\rangle_{z}\langle1|_{z}\langle-1|_{z}+|-1\rangle_{z}|1\rangle_{z}\langle-1|_{z}\langle1|_{z})$.
The colours represent approach to different states: green for the
mixed state $\rho_{{\rm st}}=\frac{1}{2}(|1\rangle_{z}|-1\rangle_{z}\langle1|_{z}\langle-1|_{z}+|-1\rangle_{z}|1\rangle_{z}\langle-1|_{z}\langle1|_{z})$,
orange the $|-1\rangle_{z}|-1\rangle_{z}$ state and blue the $|1\rangle_{z}|1\rangle_{z}$
state. Measurement strength $a=1$, duration 16, 137 trajectories.
\label{fig:ent_prod_mixed_state}}

\end{figure}

The next initial condition we consider is the mixed state $\rho_{{\rm in}}=\frac{1}{4}(|1\rangle_{z}|1\rangle_{z}\langle1|_{z}\langle1|_{z}+|-1\rangle_{z}|-1\rangle_{z}\langle-1|_{z}\langle-1|_{z}+|1\rangle_{z}|-1\rangle_{z}\langle1|_{z}\langle-1|_{z}+|-1\rangle_{z}|1\rangle_{z}\langle-1|_{z}\langle1|_{z})$
where the system has equal probabilities to collapse to the four eigenstates
of $S_{z}$: $|1\rangle_{z}|1\rangle_{z}$, $|-1\rangle_{z}|-1\rangle_{z}$,
$|\frac{1}{\sqrt{2}}(|1\rangle_{z}|-1\rangle_{z}+|-1\rangle_{z}|1\rangle_{z})$
as before and now including the singlet state $\frac{1}{\sqrt{2}}(|1\rangle_{z}|-1\rangle_{z}-|-1\rangle_{z}|1\rangle_{z})$.
As expected, the system can collapse to $|1\rangle_{z}|1\rangle_{z}$
and $|-1\rangle_{z}|-1\rangle_{z}$ each with probability $\frac{1}{4}$,
but the system is now found to evolve with a probability of $\frac{1}{2}$
to the stationary mixed state $\rho_{{\rm st}}=\frac{1}{2}(|1\rangle_{z}|-1\rangle_{z}\langle1|_{z}\langle-1|_{z}+|-1\rangle_{z}|1\rangle_{z}\langle-1|_{z}\langle1|_{z})$
which has a purity of $\frac{1}{2}$. Such a state is an equal mixture
of the two degenerate eigenstates: $\frac{1}{\sqrt{2}}(|1\rangle_{z}|-1\rangle_{z}\pm|-1\rangle_{z}|1\rangle_{z})$.
Since $S_{z}\vert j\rangle\vert k\rangle=\frac{1}{2}(j+k)\vert j\rangle\vert k\rangle$
we have $S_{z}\rho_{{\rm st}}=\rho_{{\rm st}}S_{z}=0$ which makes
$\rho_{{\rm st}}$ a stationary state of the dynamics Eq. (\ref{eq:SME-1}),
just like the eigenstates of $S_{z}$. The mean environmental stochastic
entropy productions associated with collapse to the $|1\rangle_{z}|1\rangle_{z}$
(blue), and $|-1\rangle_{z}|-1\rangle_{z}$ (orange) eigenstates as
well as the mixed state $\rho_{{\rm st}}=\frac{1}{2}(|1\rangle_{z}|-1\rangle_{z}\langle1|_{z}\langle-1|_{z}+|-1\rangle_{z}|1\rangle_{z}\langle-1|_{z}\langle1|_{z})$
(green), are illustrated in Figure \ref{fig:ent_prod_mixed_state}.

The mean asymptotic rate of environmental entropy production associated
with reaching the $|1\rangle_{z}|1\rangle_{z}$ and $|-1\rangle_{z}|-1\rangle_{z}$
eigenstates is higher than in the approach towards the mixed state
$\rho_{{\rm st}}=\frac{1}{2}(|1\rangle_{z}|-1\rangle_{z}\langle1|_{z}\langle-1|_{z}+|-1\rangle_{z}|1\rangle_{z}\langle-1|_{z}\langle1|_{z})$.
Most likely the lower rate of entropy production associated with adopting
the mixed state can be attributed to its lower purity. Furthermore,
since it lies between the $|1\rangle_{z}|1\rangle_{z}$ and $|-1\rangle_{z}|-1\rangle_{z}$
eigenstates in the $\langle S_{x}\rangle,\langle S_{z}\rangle$ coordinate
space, it could suffer from the additional competition in selection,
potentially leading to the initially slower mean rate of production.

In order to further purify the system we could measure another operator
which commutes with $S_{z}$, such as $S^{2}$, for which the $\frac{1}{\sqrt{2}}(|1\rangle_{z}|-1\rangle_{z}\pm|-1\rangle_{z}|1\rangle_{z})$
eigenstates are not degenerate. Indeed introducing an additional Lindblad
$L_{2}=S^{2}$ into the dynamics can successfully collapse the system
at one of the two $\frac{1}{\sqrt{2}}(|1\rangle_{z}|-1\rangle_{z}\pm|-1\rangle_{z}|1\rangle_{z})$
eigenstates. In contrast, introducing an additional Lindblad $L_{2}=S_{z,2}$
leads to the system purifying to $|1\rangle_{z}|-1\rangle_{z}$ or
$|-1\rangle_{z}|1\rangle_{z}$.

\subsection*{Case E.}

In the cases considered thus far, the $z$-spin components of the
first and second spin 1/2 particle have been equal such that $s_{3}=s_{12}$
throughout the measurement of $S_{z}$. Here we instead consider
an initial state: $|\Psi\rangle=\frac{1}{2}(|1\rangle_{z}|1\rangle_{z}+|-1\rangle_{z}|-1\rangle_{z})+\frac{1}{\sqrt{2}}|1\rangle_{z}|-1\rangle_{z}$,
where the $z$-spin components of particle 1 and 2 initially are:
$\frac{1}{4}$ and $-\frac{1}{4}$ respectively . As a result, $s_{3}\neq s_{12}$
throughout the measurement and unlike the previous initial states,
the particles are distinguishable and their behaviour uncorrelated.
  The system has a probability of $\frac{1}{2}$ of collapsing at
the $|1\rangle_{z}|-1\rangle_{z}$ state and a probability of $\frac{1}{4}$
of collapsing to the $|1\rangle_{z}|1\rangle_{z}$ and $|-1\rangle_{z}|-1\rangle_{z}$
eigenstates. 

\begin{figure}
\centering{}\includegraphics[width=1\columnwidth]{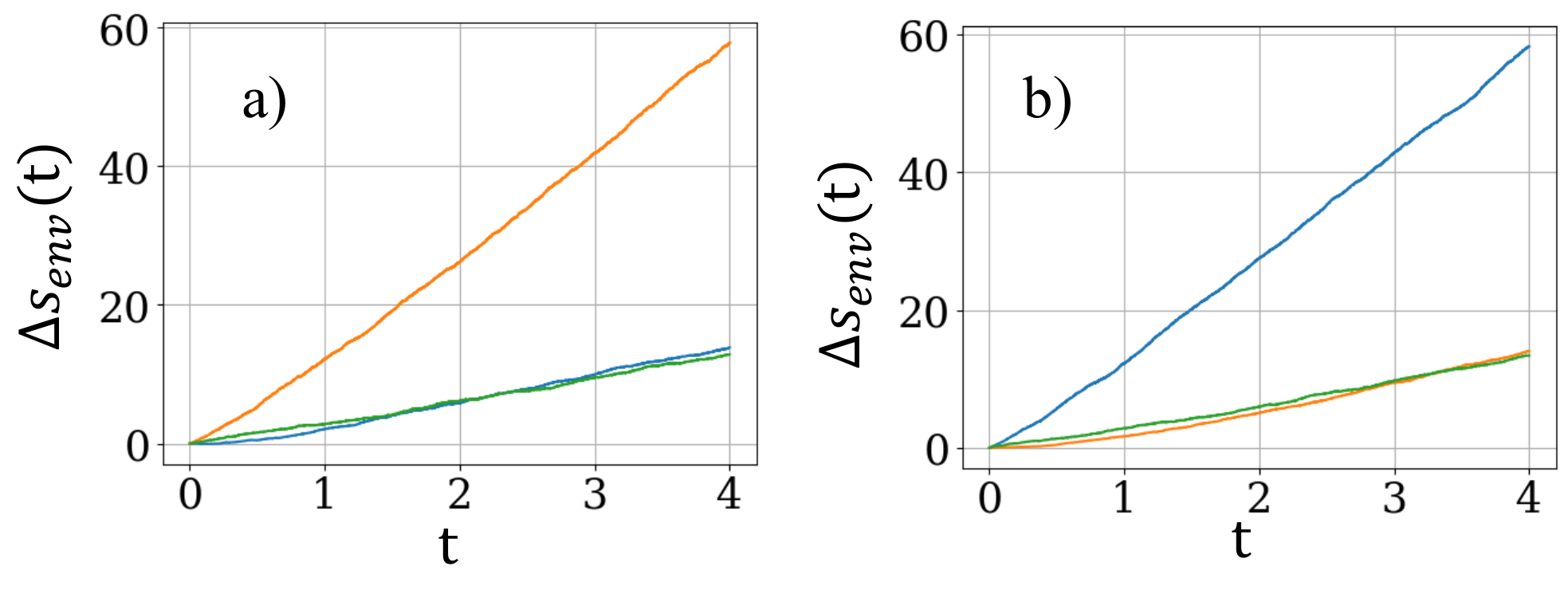}\caption{The mean environmental stochastic entropy production of the two spin
1/2 system starting in the state: $|\Psi\rangle=\frac{1}{2}(|1\rangle_{z}|1\rangle_{z}+|-1\rangle_{z}|-1\rangle_{z})+\frac{1}{\sqrt{2}}|1\rangle_{z}|-1\rangle_{z}$
for: a) $s_{3}$ as the dynamical variable.135 trajectories. b)
$s_{12}$ as the dynamical variable, 200 trajectories. Different colours
representing approach to different states for $a=1$. Orange is the
$|-1\rangle_{z}|-1\rangle_{z}$ eigenstate, blue the $|1\rangle_{z}|1\rangle_{z}$
eigenstate and green the $|1\rangle_{z}|-1\rangle_{z}$ state. \label{fig:ent_prod_eq_prob_Sz_eigens}}
\end{figure}

The mean environmental stochastic entropy production associated with
collapse to each of these eigenstates is depicted in Figure \ref{fig:ent_prod_eq_prob_Sz_eigens}
for two different choices of the dynamical variable. In Figure \ref{fig:ent_prod_eq_prob_Sz_eigens}a,
$s_{3}$ is the dynamical variable, whereas in Figure \ref{fig:ent_prod_eq_prob_Sz_eigens}b
it is $s_{12}$.

In Figure \ref{fig:ent_prod_eq_prob_Sz_eigens}a, the mean asymptotic
rate of environmental stochastic entropy production associated with
collapse to the $|-1\rangle_{z}|-1\rangle_{z}$ eigenstate (orange)
is much greater than for the $|1\rangle_{z}|-1\rangle_{z}$ state
(green) or the $|1\rangle_{z}|1\rangle_{z}$ eigenstate (blue).

To produce Figure \ref{fig:ent_prod_eq_prob_Sz_eigens}a we chose
$s_{3}$ as the dynamical variable, but if we instead pick $s_{12}$
we obtain Figure \ref{fig:ent_prod_eq_prob_Sz_eigens}b. The mean
asymptotic rates of environmental entropy production associated with
collapse to the $|-1\rangle_{z}|-1\rangle_{z}$ eigenstate and the
$|1\rangle_{z}|1\rangle_{z}$ state are now switched compared with
Figure \ref{fig:ent_prod_eq_prob_Sz_eigens}a.

\begin{figure}
\begin{centering}
\includegraphics[width=1\columnwidth]{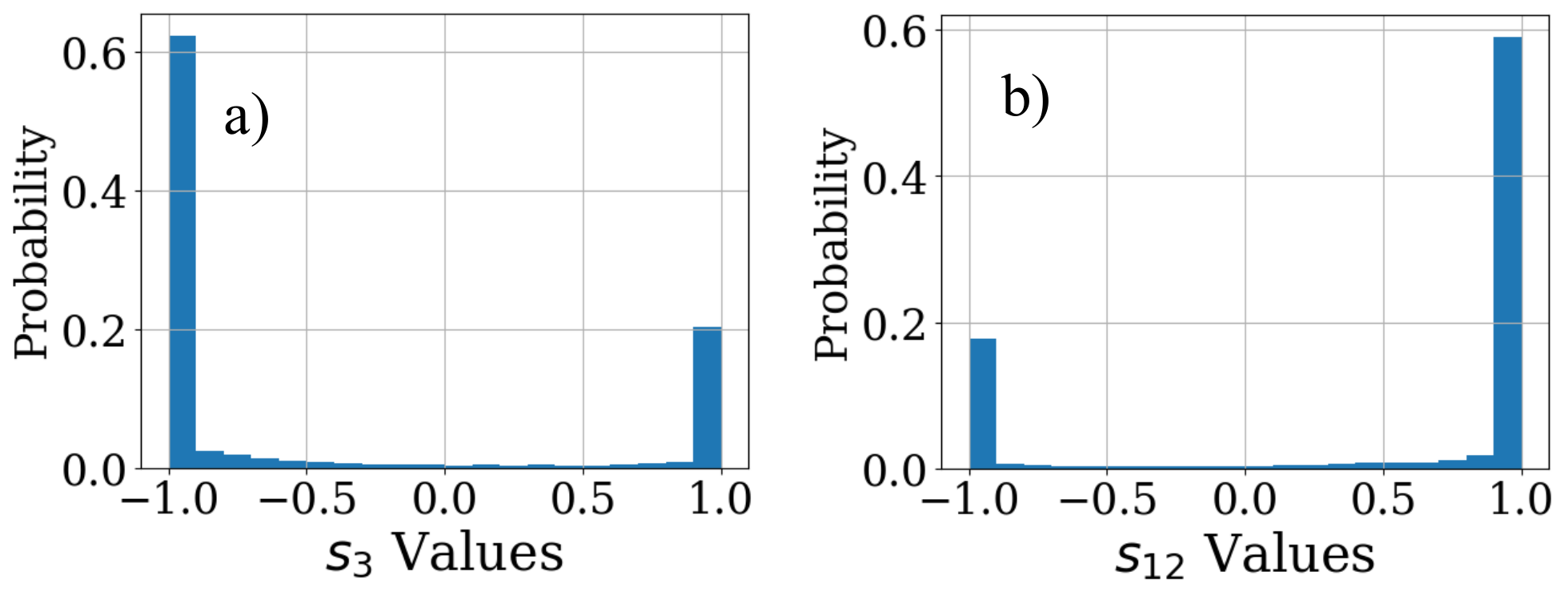}
\par\end{centering}
\caption{The probability density functions of the dynamical variables (a) $s_{3}$
and (b) $s_{12}$ after an $S_{z}$ measurement performed on the two
spin 1/2 particles in Case E. Measurement strength $a=1$, measurement
duration: 12. \label{fig:pdfs_s3_s12}}
\end{figure}

At first sight this dependence is unexpected, but it can be resolved
as follows. The total entropy production $\Delta s_{{\rm tot}}=\Delta s_{{\rm env}}+\Delta s_{{\rm sys}}$
does not depend on the choice of dynamical variable but $\Delta s_{{\rm env}}$
and $\Delta s_{{\rm sys}}$ can do so. In the case studied, choosing
$s_{3}$ or $s_{12}$ as dynamical variable changes the asymptotic
rate of environmental entropy production associated with approach
to the $|-1\rangle_{z}|-1\rangle_{z}$ eigenstate or the $|1\rangle_{z}|1\rangle_{z}$
eigenstate, and we would therefore expect the system stochastic entropy
production to reflect this difference.

Recall that the system stochastic entropy production when $s_{3}$
is chosen as the dynamical variable evolves as $d\Delta s_{{\rm sys}}=-d\text{ln}p(s_{3},t)$,
where $p(s_{3},t)$ is the probability density function (pdf) of $s_{3}$,
and similar for $s_{12}$. Pdfs for both $s_{3}$ and $s_{12}$ after
a period of measurement are shown in Figure \ref{fig:pdfs_s3_s12}.
The initial state is such that the system has a probability of $\frac{1}{2}$
of collapsing to the $|1\rangle_{z}|-1\rangle_{z}$ eigenstate and
a probability of $\frac{1}{4}$ of collapsing to the $|1\rangle_{z}|1\rangle_{z}$
and $|-1\rangle_{z}|-1\rangle_{z}$ eigenstates. Therefore $s_{12}$,
has a probability of $\frac{3}{4}$ of collapsing at $+1$ and $\frac{1}{4}$
of collapsing at $-1$, whilst $s_{3}$, has a probability of $\frac{3}{4}$
of collapsing at $-1$ and $\frac{1}{4}$ of collapsing at $+1$,
i.e. the other way round. Since the system will not always have enough
time to collapse precisely at an eigenstate, the pdfs in Figure \ref{fig:pdfs_s3_s12}
are somwhat broader.

Nonetheless, from Figure \ref{fig:pdfs_s3_s12}, we can infer that
 $-d\text{ln}p(s_{3},t)\neq-d\text{ln}p(s_{12},t)$ and the system
entropy production is different according to whether $s_{3}$ or $s_{12}$
is picked to be the dynamical variable. The asymmetry between the
pdfs of the $z$-spins of the two spin 1/2 particles, $p(s_{3},t)\neq p(s_{12},t)$,
in this Case is the source of the dependence of the environmental
stochastic entropy production on the choice of the dynamical variable.

\section{Conclusions\label{sec:Conclusions}}

Calculating environmental stochastic entropy production associated
with quantum measurement is problematic in the conventional framework
of discontinuous and instantaneous wavefunction collapse. In contrast,
it is natural in the quantum state diffusion (QSD) formalism where
measurement collapse is continuous and takes place over a finite period
of time. Measurement is brought about by an interaction between a
system and an environment (the measurement apparatus), causing the
system to diffuse towards an eigenstate of the observable in question.
The speed of collapse is related to the coupling strength between
the system and the environment.

Using stochastic quantum trajectories generated in such a framework
we calculate the environmental stochastic entropy production associated
with measurement of a system of two spin 1/2 particles, selected according
to initial and final states and measured property. In particular we
compare the mean asymptotic rates of environmental stochastic entropy
production when the system is close to an eigenstate. Such entropy
production differs from changes in the von Neumann entropy in that
it applies to single realisations of system behaviour as opposed to
average evolution and concerns the irreversibility of the approach
towards a measurement outcome, as opposed to uncertainty in the choice
of the measurement outcomes themselves. Whilst the von Neumann entropy
takes an upper limit of $\text{ln}2$, the environmental stochastic
entropy production has no upper limit.

The type of measurement performed appears to affect the mean asymptotic
rate of environmental stochastic entropy production. In Case 1 single
particle $z$-spin measurements are performed on both particles, and
in Case 2 single particle $z$-spin and $x$-spin measurements are
performed on the first and second particles respectively. The rate
of production is almost two times greater in the latter case, as shown
in Figures \ref{fig:z_z_mes_ent} and \ref{fig:z_x_ent_prod}. 

Additionally, the mean asymptotic rate associated with collapse to
the same eigenstate appears to depend on the initial state of the
system. For example, in a situation where the total $z$-spin $S_{z}$
of the two particle system is measured, the mean asymptotic production
associated with collapse to $|1\rangle_{z}|1\rangle_{z}$ is four
times greater for Case A (Figure \ref{fig:2_spin_half_entropy_start_Sx_0})
than Case B (Figure \ref{fig:2_spin_half_ent_prod_eq_sup_sz_tripzero_sz_1}).
The system was prepared in different states but each with a probability
of $\frac{1}{2}$ of collapsing to $|1\rangle_{z}|1\rangle_{z}$.
The same can be said for Cases B and C where collapse to the $\frac{1}{\sqrt{2}}(|1\rangle_{z}|-1\rangle_{z}+|-1\rangle_{z}|1\rangle_{z})$
eigenstate is associated with different mean asymptotic rates of production
in Figures \ref{fig:2_spin_half_ent_prod_eq_sup_sz_tripzero_sz_1}
and \ref{fig:The-average-stochastic2_spin_half_entropy_start_Sx_1}.
Again the system starts in different states but with the same probability
of collapsing at $\frac{1}{\sqrt{2}}(|1\rangle_{z}|-1\rangle_{z}+|-1\rangle_{z}|1\rangle_{z})$.

Case C demonstrates that an approach towards eigenstates with the
highest and lowest eigenvalues appears to exhibit a higher asymptotic
rate of production than collapse to the eigenstate in the middle,
as shown in Figures \ref{fig:The-average-stochastic2_spin_half_entropy_start_Sx_1}
and \ref{fig:ent_prod_mixed_state}. This appears to be the case only
when three or more eigenstates are involved since this distinction
does not present itself in Case B (see Figure \ref{fig:2_spin_half_ent_prod_eq_sup_sz_tripzero_sz_1}).

The way in which QSD handles degenerate eigenstates was demonstrated
by Case D, where the system `collapsed' to a state which was not pure
but nevertheless a stationary state under the dynamics. Only by additionally
measuring $S^{2}$ or a single particle $z$-spin operator did the
system finally purify. 

Finally, Case E illustrates an instance where the environmental stochastic
entropy production depends on the choice of the dynamical variable.
Such differences are reflected also in the system stochastic entropy
production, thus, as we would expect, the total entropy production
remains invariant to the choice of dynamical variable. The initial
state brought about an asymmetry in the pdfs of the two spin 1/2 particle
$z$-spin components, unlike the other Cases considered.

The environmental stochastic entropy production offers an additional
way of characterising measurement collapse that may uncover new insights
into the process, and we have presented various examples in this work.
One additional avenue to explore is to use environmental stochastic
entropy production to distinguish two forms of mixed density matrices:
one representing a single system entangled with its environment, and
one that merely represents an ensemble of pure quantum states. When
the system is already in an eigenstate of the measured operator, it
remains stationary and therefore its environmental stochastic entropy
production under measurement would be zero. If however the system
is in an entangled state it will evolve towards an eigenstate and
the environmental stochastic entropy production would be non-zero.
The production of this quantity, and the irreversibility under measurement
that it represents, can distinguish situations that are typically
regarded as indistinguishable.

\section*{Acknowledgements}

SMW is supported by a PhD studentship funded by the Engineering and
Physical Sciences Research Council (EPSRC) under grant codes EP/R513143/1
and EP/T517793/1.

\bibliography{Version_7/entropy_prod_paper_draft_7}

\appendix

\section{General SDE for stochastic entropy production \label{sec:Appendix-A}}

We consider a set of coordinates $\textbf{\ensuremath{\boldsymbol{x}}}=\{x_{1},x_{2},...,x_{N}\}$
whose dynamics are defined by It\^{o} processes:

\begin{equation}
dx_{i}=A_{i}(\boldsymbol{x})dt+\sum_{j}B_{ij}(\boldsymbol{x})dW_{j},\label{eq:ito_sde}
\end{equation}
where $dW_{j}$ denote independent Wiener increments. $A_{i}(\boldsymbol{x})$
may be split into reversible $A_{i}^{{\rm rev}}(\boldsymbol{x})$
and irreversible $A_{i}^{{\rm irr}}(\boldsymbol{x})$ contributions
as follows \citep{spinney2012,dexter2025}:

\begin{align}
A_{i}^{{\rm irr}}(\boldsymbol{x}) & =\frac{1}{2}\left[A_{i}(\boldsymbol{x})+\epsilon_{i}A_{i}(\epsilon\boldsymbol{x})\right]=\epsilon_{i}A_{i}^{{\rm irr}}(\epsilon\boldsymbol{x})\nonumber \\
A_{i}^{{\rm rev}}(\boldsymbol{x}) & =\frac{1}{2}\left[A_{i}(\boldsymbol{x})-\epsilon_{i}A_{i}(\epsilon\boldsymbol{x})\right]=-\epsilon_{i}A_{i}^{{\rm rev}}(\epsilon\boldsymbol{x}),
\end{align}
where $\epsilon_{i}=1$ for variables $x_{i}$ with even parity under
time reversal symmetry, whilst $\epsilon_{i}=-1$ for variables with
odd parity. The form $\boldsymbol{\varepsilon}\boldsymbol{x}$ represents
$(\varepsilon_{1}x_{1},\varepsilon_{2}x_{2},\cdots)$. As the notation
suggests, the $A_{i}^{{\rm irr}}$ terms are partially responsible
for irreversible behaviour while the $A_{i}^{{\rm rev}}$ are time
reversal invariant and are not responsible. Defining an $N\times N$
diffusion matrix $\boldsymbol{D}=\frac{1}{2}\boldsymbol{B}\boldsymbol{B}^{\mathsf{T}}$,
the evolution of the environmental stochastic entropy production is
then governed by \citep{spinney2012}:

\begin{equation}
\begin{split} & d\Delta s_{{\rm env}}=-\sum_{i}\frac{\partial A_{i}^{\text{rev}}}{\partial x_{i}}dt+\sum_{i,j}\Biggl\{ D_{ij}^{-1}A_{i}^{\text{irr}}dx_{j}\\
 & -D_{ij}^{-1}\sum_{m}\frac{\partial D_{im}}{\partial x_{m}}dx_{j}-D_{ij}^{-1}A_{i}^{\text{rev}}A_{j}^{\text{irr}}dt\\
 & +D_{ij}^{-1}A_{i}^{\text{rev}}\sum_{n}\frac{\partial D_{jn}}{\partial x_{n}}dt+\sum_{k}\Biggl[D_{ik}\frac{\partial}{\partial x_{k}}\left(D_{ij}^{-1}A_{j}^{\text{irr}}\right)\\
 & -D_{ik}\frac{\partial}{\partial x_{k}}\left(D_{ij}^{-1}\sum_{n}\frac{\partial D_{jn}}{\partial x_{n}}\right)\Biggr]dt\Biggr\}.
\end{split}
\label{eq: senvbig_long}
\end{equation}

For a system described by even parity variables with $\varepsilon_{i}=1$,
the reversible terms $A_{i}^{{\rm rev}}$ are zero and $A_{i}^{{\rm irr}}=A_{i}$.
Eq. (\ref{eq: senvbig_long}) then simplifies to
\begin{equation}
\begin{split} & d\Delta s_{{\rm env}}=\sum_{i,j}\Biggl\{ D_{ij}^{-1}A_{i}dx_{j}-D_{ij}^{-1}\sum_{m}\frac{\partial D_{im}}{\partial x_{m}}dx_{j}+\\
 & \sum_{k}\Biggl[D_{ik}\frac{\partial}{\partial x_{k}}\left(D_{ij}^{-1}A_{j}\right)-D_{ik}\frac{\partial}{\partial x_{k}}\left(D_{ij}^{-1}\sum_{m}\frac{\partial D_{jm}}{\partial x_{m}}\right)\Biggr]dt\Biggr\}.
\end{split}
\label{eq: senvbig_long-2}
\end{equation}
which then takes the compact form of Eq. (\ref{eq: senvbig}).

\section{Singular diffusion matrices\label{sec:Appendix-B}}

Singularity in the diffusion matrix may be attributed to the existence
of functions of the stochastic variables that either evolve deterministically
or are constants of the motion. This implies that the irreversibility
of the motion is captured by a subset of what we call \emph{dynamical}
variables. It is possible to establish a relationship between the
dynamical and remaining \emph{spectator} variables for the purposes
of calculating an entropy production \citep{dexter2025}. More specifically,
the derivatives of the diffusion matrix elements in Eq. (\ref{eq: senvbig_long-2}),
can be recast into a more complex expression.

We can define the relationship as 
\begin{equation}
\sum_{l=1}^{L}\alpha_{kl}dx_{l}+\sum_{m=1}^{M}\alpha_{km}dx_{m}=G_{k}dt,\label{eq:29}
\end{equation}
where $\alpha_{km}$ and $\alpha_{kl}$ are the dynamical and spectator
variable components of the $k^{th}$ null eigenvector of the diffusion
matrix (an eigenvector with an eigenvalue of zero) and $G_{k}$ is
a function of the stochastic variables whose form we need not discuss
\citep{dexter2025}. We write
\begin{equation}
dx_{l}=-P_{lk}^{-1}Q_{km}dx_{m}+P_{lk}^{-1}G_{k}dt=R_{lm}dx_{m}+S_{l}dt,\label{eq:30}
\end{equation}
where the $\alpha_{km}$ have been arranged as elements of a rectangular
$L\times M$ matrix $Q_{km}$ and the $\alpha_{kl}$ as elements of
a square $L\times L$ matrix $P_{kl}$. The $L\times M$ matrix $\boldsymbol{R}$
is defined as $\boldsymbol{R}=-\boldsymbol{P}^{-1}\boldsymbol{Q}$.

The derivatives $dD_{ij}/dx_{m}$ in Eq. (\ref{eq: senvbig_long-2})
are then given by a combination of derivatives with respect to dynamical
variables $x_{m}$ and the spectator variables $x_{l}$, corresponding
to Eq. (\ref{eq:corr_terms}) in the main text \citep{dexter2025}:

\begin{equation}
\frac{dD_{ij}}{dx_{m}}=\frac{\partial D_{ij}}{\partial x_{m}}+\sum_{l}\frac{\partial D_{ij}}{\partial x_{l}}\frac{dx_{l}}{dx_{m}}=\frac{\partial D_{ij}}{\partial x_{m}}+\sum_{l}\frac{\partial D_{ij}}{\partial x_{l}}R_{lm}.\label{eq:corr_terms2-1}
\end{equation}

\pagebreak{}

\section{Generator matrices for a two spin-1/2 particle system \label{sec:Appendix-C}}

The SU(2)$\otimes$SU(2) generators used to form the entangled two
spin 1/2 system density matrix in Eq. (\ref{eq:rho_ent_spins}) are
as follows: $\Sigma_{1}=I_{2}\otimes\sigma_{x}$, $\Sigma_{2}=I_{2}\otimes\sigma_{y}$,
$\Sigma_{3}=I_{2}\otimes\sigma_{z}$, $\Sigma_{4}=\sigma_{x}\otimes I_{2}$,
$\Sigma_{5}=\sigma_{x}\otimes\sigma_{x}$, $\Sigma_{6}=\sigma_{x}\otimes\sigma_{y}$,
$\Sigma_{7}=\sigma_{x}\otimes\sigma_{z}$, $\Sigma_{8}=\sigma_{y}\otimes I_{2}$,
$\Sigma_{9}=\sigma_{y}\otimes\sigma_{x}$, $\Sigma_{10}=\sigma_{y}\otimes\sigma_{y}$,
$\Sigma_{11}=\sigma_{y}\otimes\sigma_{z}$, $\Sigma_{12}=\sigma_{z}\otimes I_{2}$,
$\Sigma_{13}=\sigma_{z}\otimes\sigma_{x}$, $\Sigma_{14}=\sigma_{z}\otimes\sigma_{y}$
and $\Sigma_{15}=\sigma_{z}\otimes\sigma_{z}$, where $I_{2}$ is
a $2\times2$ identity matrix. Explicitly:

\begin{align}
 & \Sigma_{1}=\begin{pmatrix}0 & 1 & 0 & 0\\
1 & 0 & 0 & 0\\
0 & 0 & 0 & 1\\
0 & 0 & 1 & 0
\end{pmatrix}\,\quad\Sigma_{2}=\begin{pmatrix}0 & -i & 0 & 0\\
i & 0 & 0 & 0\\
0 & 0 & 0 & -i\\
0 & 0 & i & 0
\end{pmatrix}\nonumber \\
 & \Sigma_{3}=\begin{pmatrix}1 & 0 & 0 & 0\\
0 & -1 & 0 & 0\\
0 & 0 & 1 & 0\\
0 & 0 & 0 & -1
\end{pmatrix}\,\quad\Sigma_{4}=\begin{pmatrix}0 & 0 & 1 & 0\\
0 & 0 & 0 & 1\\
1 & 0 & 0 & 0\\
0 & 1 & 0 & 0
\end{pmatrix}\nonumber \\
 & \Sigma_{5}=\begin{pmatrix}0 & 0 & 0 & 1\\
0 & 0 & 1 & 0\\
0 & 1 & 0 & 0\\
1 & 0 & 0 & 0
\end{pmatrix}\,\quad\Sigma_{6}=\begin{pmatrix}0 & 0 & 0 & -i\\
0 & 0 & i & 0\\
0 & -i & 0 & 0\\
i & 0 & 0 & 0
\end{pmatrix}\nonumber \\
 & \Sigma_{7}=\begin{pmatrix}0 & 0 & 1 & 0\\
0 & 0 & 0 & -1\\
1 & 0 & 0 & 0\\
0 & -1 & 0 & 0
\end{pmatrix}\,\quad\Sigma_{8}=\begin{pmatrix}0 & 0 & -i & 0\\
0 & 0 & 0 & -i\\
i & 0 & 0 & 0\\
0 & i & 0 & 0
\end{pmatrix}\nonumber \\
 & \Sigma_{9}=\begin{pmatrix}0 & 0 & 0 & -i\\
0 & 0 & -i & 0\\
0 & i & 0 & 0\\
i & 0 & 0 & 0
\end{pmatrix}\,\quad\Sigma_{10}=\begin{pmatrix}0 & 0 & 0 & -1\\
0 & 0 & 1 & 0\\
0 & -1 & 0 & 0\\
1 & 0 & 0 & 0
\end{pmatrix}\nonumber \\
 & \Sigma_{11}=\begin{pmatrix}0 & 0 & -i & 0\\
0 & 0 & 0 & i\\
i & 0 & 0 & 0\\
0 & -i & 0 & 0
\end{pmatrix}\,\quad\Sigma_{12}=\begin{pmatrix}1 & 0 & 0 & 0\\
0 & 1 & 0 & 0\\
0 & 0 & -1 & 0\\
0 & 0 & 0 & -1
\end{pmatrix}\nonumber \\
 & \Sigma_{13}=\begin{pmatrix}0 & 1 & 0 & 0\\
1 & 0 & 0 & 0\\
0 & 0 & 0 & -1\\
0 & 0 & -1 & 0
\end{pmatrix}\,\quad\Sigma_{14}=\begin{pmatrix}0 & -i & 0 & 0\\
i & 0 & 0 & 0\\
0 & 0 & 0 & i\\
0 & 0 & -i & 0
\end{pmatrix}\nonumber \\
 & \Sigma_{15}=\begin{pmatrix}1 & 0 & 0 & 0\\
0 & -1 & 0 & 0\\
0 & 0 & -1 & 0\\
0 & 0 & 0 & 1
\end{pmatrix}.\label{eq:su2_generators-1}
\end{align}
\pagebreak{}

\section{SDEs for Case 1\label{sec:Appendix-D}}

The It\^{o} processes for the variables parametrising the two spin
1/2 system undergoing $z$-spin measurements on each particle are
the following, for $a_{1}=a_{2}=1$:

\begin{equation}
\begin{aligned}ds_{1} & =-\frac{1}{2}s_{1}dt+(-s_{1}s_{12}+s_{13})dW_{1}-s_{1}s_{3}dW_{2}\\
ds_{2} & =-\frac{1}{2}s_{2}dt+(-s_{2}s_{12}+s_{14})dW_{1}-s_{2}s_{3}dW_{2}\\
ds_{3} & =(-s_{3}s_{12}+s_{15})dW_{1}+(1-s_{3}^{2})dW_{2}\\
ds_{4} & =-\frac{1}{2}s_{4}dt-s_{4}s_{12}dW_{1}+(-s_{3}s_{4}+s_{7})dW_{2}\\
ds_{5} & =-s_{5}dt-s_{5}s_{12}dW_{1}+(-s_{3}s_{5}+s_{4})dW_{2}\\
ds_{6} & =-s_{6}dt-s_{6}s_{12}dW_{1}-s_{3}s_{6}dW_{2}\\
ds_{7} & =-\frac{1}{2}s_{7}dt-s_{7}s_{12}dW_{1}-(s_{4}-s_{3}s_{7})dW_{2}\\
ds_{8} & =-\frac{1}{2}s_{8}dt-s_{8}s_{12}dW_{1}+(-s_{3}s_{8}+s_{11})dW_{2}\\
ds_{9} & =-s_{9}dt-s_{9}s_{12}dW_{1}-s_{3}s_{9}dW_{2}\\
ds_{10} & =-s_{10}dt-s_{10}s_{12}dW_{1}-s_{3}s_{10}dW_{2}\\
ds_{11} & =-\frac{1}{2}s_{11}dt-s_{11}s_{12}dW_{1}+(s_{8}-s_{3}s_{11})dW_{2}\\
ds_{12} & =(1-s_{12}^{2})dW_{1}+(-s_{3}s_{12}+s_{15})dW_{2}\\
ds_{13} & =-\frac{1}{2}s_{13}dt+(s_{1}-s_{12}s_{13})dW_{1}-s_{3}s_{13}dW_{2}\\
ds_{14} & =-\frac{1}{2}s_{14}dt+(s_{2}-s_{12}s_{14})dW_{1}-s_{3}s_{14}dW_{2}\\
ds_{15} & =(s_{3}-s_{12}s_{15})dW_{1}+(s_{12}-s_{3}s_{15})dW_{2}.
\end{aligned}
\end{equation}
\pagebreak{}

\section{SDEs for Case 2 \label{sec:Appendix-E}}

The It\^{o} processes for the variables parametrising the two spin
1/2 system undergoing a $z$-spin measurement on particle 1 and an
$x$-spin measurement on particle 2 are the following, for $a_{1}=a_{2}=1$:

\begin{equation}
\begin{aligned}ds_{1} & =(-s_{1}s_{12}+s_{13})dW_{1}+(1-s_{1}^{2})dW_{2}\\
ds_{2} & =-\frac{1}{2}s_{2}dt+(-s_{2}s_{12}+s_{14})dW_{1}-s_{1}s_{2}dW_{2}\\
ds_{3} & =-\frac{1}{2}s_{3}dt+(-s_{3}s_{12}+s_{15})dW_{1}-s_{1}s_{3}dW_{2}\\
ds_{4} & =-\frac{1}{2}s_{4}dt-s_{4}s_{12}dW_{1}+(-s_{1}s_{4}+s_{5})dW_{2}\\
ds_{5} & =-\frac{1}{2}s_{5}dt-s_{5}s_{12}dW_{1}+(-s_{1}s_{5}+s_{4})dW_{2}\\
ds_{6} & =-s_{6}dt-s_{6}s_{12}dW_{1}-s_{1}s_{6}dW_{2}\\
ds_{7} & =-s_{7}dt-s_{7}s_{12}dW_{1}-s_{1}s_{7}dW_{2}\\
ds_{8} & =-\frac{1}{2}s_{8}dt-s_{8}s_{12}dW_{1}+(-s_{1}s_{8}+s_{9})dW_{2}\\
ds_{9} & =-\frac{1}{2}s_{9}dt-s_{9}s_{12}dW_{1}+(-s_{1}s_{9}+s_{8})dW_{2}\\
ds_{10} & =-s_{10}dt-s_{10}s_{12}dW_{1}-s_{1}s_{10}dW_{2}\\
ds_{11} & =-s_{11}dt-s_{11}s_{12}dW_{1}-s_{1}s_{11}dW_{2}\\
ds_{12} & =(1-s_{12}^{2})dW_{1}+(-s_{1}s_{12}+s_{13})dW_{2}\\
ds_{13} & =(s_{1}-s_{12}s_{13})dW_{1}+(s_{12}-s_{1}s_{13})dW_{2}\\
ds_{14} & =-\frac{1}{2}s_{14}dt+(s_{2}-s_{12}s_{14})dW_{1}-s_{1}s_{14}dW_{2}\\
ds_{15} & =-\frac{1}{2}s_{15}dt+(s_{3}-s_{12}s_{15})dW_{1}-s_{1}s_{15}dW_{2}.
\end{aligned}
\end{equation}
\pagebreak{}

\section{SDEs for measurement of total $z$ component of spin \label{sec:Appendix-G}}

It\^{o} processes for the variables parametrising the two spin 1/2
system density matrix when the system is undergoing a measurement
of total $S_{z}$ are as follows: 

\begin{align}
ds_{1} & =-\frac{1}{2}a^{2}s_{1}dt+a(s_{13}-s_{1}(s_{3}+s_{12}))dW_{z}\nonumber \\
ds_{2} & =-\frac{1}{2}a^{2}s_{2}dt+a(s_{14}-s_{2}(s_{3}+s_{12}))dW_{z}\nonumber \\
ds_{3} & =a(1-s_{3}(s_{3}+s_{12})+s_{15})dW_{z}\nonumber \\
ds_{4} & =\frac{1}{2}a^{2}s_{4}dt+a(s_{7}-s_{4}(s_{3}+s_{12}))dW_{z}\nonumber \\
ds_{5} & =a^{2}(s_{10}-s_{5})dt-as_{5}(s_{3}+s_{12})dW_{z}\nonumber \\
ds_{6} & =a^{2}(-s_{6}-s_{9})dt-as_{6}(s_{3}+s_{12})dW_{z}\nonumber \\
ds_{7} & =-\frac{1}{2}a^{2}s_{7}dt+a(s_{4}-s_{7}(s_{3}+s_{12}))dW_{z}\nonumber \\
ds_{8} & =-\frac{1}{2}a^{2}s_{8}dt+a(s_{11}-s_{8}(s_{3}+s_{12}))dW_{z}\nonumber \\
ds_{9} & =-a^{2}(-s_{6}-s_{9})dt-as_{9}(s_{3}+s_{12})dW_{z}\nonumber \\
ds_{10} & =a^{2}(s_{5}-s_{10})dt-as_{10}(s_{3}+s_{12})dW_{z}\nonumber \\
ds_{11} & =-\frac{1}{2}a^{2}s_{11}dt+a(s_{8}-s_{11}(s_{3}+s_{12}))dW_{z}\nonumber \\
ds_{12} & =a(1-s_{12}(s_{3}+s_{12})+s_{15})dW_{z}\nonumber \\
ds_{13} & =-\frac{1}{2}a^{2}s_{13}dt-a(s_{3}s_{13}+s_{12}s_{13}-s_{1})dW_{z}\nonumber \\
ds_{14} & =-\frac{1}{2}a^{2}s_{14}dt+a(s_{2}-s_{14}(s_{3}+s_{12}))dW_{z}\nonumber \\
ds_{15} & =-a(s_{3}+s_{12})(-1+s_{15})dW_{z}.\label{SDEs for ent spin}
\end{align}

\end{document}